\documentclass[letter]{emulateapj} \usepackage{apjfonts,psfig,epsfig}

\usepackage{amsmath, amsthm, amssymb} \usepackage{lscape}
\usepackage{longtable}

\def\lsim{\mathrel{\rlap{\lower4pt\hbox{\hskip1pt$\sim$}}
    \raise1pt\hbox{$<$}}}                
\def\gsim{\mathrel{\rlap{\lower4pt\hbox{\hskip1pt$\sim$}}
    \raise1pt\hbox{$>$}}}                

\shorttitle{} \shortauthors{Stark, Loeb, \& Ellis}

\begin{document}

\title{An Empirically-Calibrated Model For Interpreting the Evolution
  of Galaxies  During the Reionization Era}


\author {Daniel P. Stark\altaffilmark{1},  Abraham
  Loeb\altaffilmark{2}, Richard S. Ellis\altaffilmark{1} } 

\altaffiltext{1}{Department of Astrophysics, California Institute of
  Technology,  MS 105-24, Pasadena, CA 91125; dps@astro.caltech.edu}
\altaffiltext{2}{Astronomy Department, Harvard University, 60  Garden
  Street, Cambridge, Massachusetts 02138, USA }

\begin{abstract}

We develop a simple star formation model whose goal is to interpret
the emerging body of observational data on star-forming galaxies at
$z\gtrsim 6$. The efficiency and duty cycle of the star formation
activity within dark matter halos are determined by fitting the
luminosity functions of Ly$\alpha$ emitter and Lyman-break galaxies at
redshifts $z\simeq 5-6$.  Our error budget takes proper account of the
uncertainty arising from both the spatial clustering of galaxies and
the commonly-used Poisson contribution.  For a given survey geometry,
we find a cross-over luminosity below which clustering provides the
dominant contribution to this variance.  Using our model parameters we
predict the likely abundance of star forming galaxies at earlier
epochs and compare these to the emerging data in the redshift interval
$7<z<10$. We find that the abundance of luminous Lyman-break galaxies
in the 500 Myr between $z\simeq 6$ and 10 can be naturally explained
by the hierarchical assembly of dark matter haloes; there is only
marginal evidence for strong physical evolution. In contrast, the
first estimates of the abundance of less luminous star forming
galaxies at $z=9-10$ are higher than predicted and, if verified by
further data, may suggest a top-heavy stellar mass function at these
early epochs. Although these abundances remain uncertain because of
the difficulty of spectroscopic confirmation and cosmic variance, even
a modest improvement in survey capability with present or upcoming
facilities should yield great progress.  In this context, we use our
model to consider those observational techniques that hold the most
promise and make predictions for specific surveys that are, or will
soon be, underway.  We conclude that narrowband Ly$\alpha$ emitter
surveys should be efficient on searches at $z\simeq 7-8$; however,
such conventional surveys are unlikely to detect sufficient galaxies
at $z\simeq 10$ to provide useful constraints.  For this reason,
gravitational  lensing offers the best prospect for probing the
$z\simeq 10$ universe  prior to JWST.

\end{abstract}


\section{Introduction}

The recent discovery of star-forming galaxies at high redshifts,
$z>6$, represents an emerging frontier in our understanding of the
early stages of galaxy formation. Such studies aim to determine the
role that young galaxies may play in completing cosmic reionization,
as well as to define more clearly how feedback and other
poorly-understood processes shape the early distribution of galaxy
luminosities and sizes from which later systems evolve (see reviews by
\citealt{Loeb2007, Ellis2007}).
 
Considerable observational progress is being made through ambitious
campaigns undertaken with the Hubble and Spitzer Space Telescopes and
large ground-based telescopes.  As a result, luminosity functions and
stellar mass distributions are now available for various star-forming
populations observed at $z\simeq 5-6$. These include the continuum
`drop-outs' or Lyman break galaxies
(LBGs,\citealt{Bou06,Yan2006,Eyles2006,Stark07a}),  and the
Lyman-alpha emitters (LAEs, \citealt{Santos04,Malhotra2004,
  Hu2004,Shimasaku06,Kashikawa06}), located either spectroscopically
or via narrow band imaging. Alongside these achievements, the first
constraints are now emerging on the abundance of equivalent systems at
$7<z<10$  \citep{Richard06,Willis2005,Bou06b,Iye,Cuby2006,Stark07b}.

Several questions arise as observers continue to make progress. First,
what is the physical relationship between the Lyman Break Galaxy (LBG)
and  Lyman-alpha emitter (LAE) populations?  This is important in
interpreting the quite significant differences that have been observed
between the properties of the two classes. Second, what redshift
trends are expected for these populations?  Some authors
(e.g. \citealt{Kashikawa06,Bou06b}) have claimed strong evolution in
the abundances with redshift.  However, in the absence of any
theoretical framework it is difficult to assess the significance of
these claims. Finally, given the abundance of star forming galaxies at
$z\simeq 5-6$, what is expected at $z\simeq 7-10$, the current
observational target? And what is the optimum observational strategy
for finding those sources which will be valuable in constraining the
epoch of cosmic reionization and the properties of young galaxies?

The present paper is motivated by the need to answer these questions
given the improving observational situation at $z\simeq$5--6. As the
time interval between $z$=6 and $z$=10 is short by cosmic standards
($\simeq$470 Myr in the WMAP3 cosmology, \citealt{Spergel2006}), the
growth of the halo mass function over this redshift range is well
understood. Accordingly, it is practical to fit the $z\simeq 5-6$
observations in the context of a simple star-formation model, thereby
deducing the likely differences between the LBG and LAE populations,
and to then use such an empirically calibrated model to predict their
likely abundances at $z\simeq 10$.

The model we adopt assumes all early star-forming galaxies observed
during this relatively short period are primarily triggered into
action by halo mergers. The key parameters governing their visibility
are thus the efficiency of star formation and the duty cycle of its
activity. An observed luminosity function at $z\simeq 5-6$ thus
provides a joint constraint on the star formation efficiency (which
determines the rate at which baryons are converted into stars) and the
duty cycle of activity (which determines the fraction of halos
occupied by visible galaxies). We consider such a simple model to be
complementary and perhaps more intuitive than full {\it ab initio}
numerical simulations which must also assume sub-grid presriptions for
star formation (e.g \citealt{Nagamine05,Gnedin06,Finlator06}).  The
goal is to infer the likely redshift trends in the context of emerging
data and to use the model to evaluate the future observational
prospects,  particularly in the optimal design of surveys to locate
sources at  $z\simeq 10$ or so.

Several authors have already made good progress with such
semi-analytic models. \cite{LeDelliou2005} and \cite{Dijkstra2006}
have attempted to fit the luminosity distribution of LAEs. Dijkstra et
al. argue that the evolution observed by \cite{Kashikawa06} between
$z=$5.7 and $z$=6.5, claimed as arising from changes in the
intergalactic medium, may be understood instead through simple growth
in the halo mass function.   \cite{Mao2006} have extended the method
to contrast the properties of LAEs and LBGs; they find LBGs reside in
a wide range of halo masses (10$^{10}$ to 10$^{12}~\rm{M_{\odot}}$),
whereas LAEs reside within a narrower range ($\simeq10^{11}
\rm{M_{\odot}}$).  A key inference from their model is the short duty
cycle of activity in the most luminous sources.  \cite{Samui06} have
argued that the evolution observed at $6\lsim z \lsim 10$ in the
Hubble Ultra  Deep Field (UDF) can be attributed to evolution in the
underlying dark matter halo number density without requiring a
dramatic change in the nature of star formation; in contrast, they
find that the large abundance of $z\simeq 9$ LBGs discovered in the
gravitational lensing survey of \cite{Richard06} requires significant
evolution in the stellar initial mass function, the reddening
correction,  and the mode of the star formation.

The present paper continues the earlier work.  We focus not only on
explaining the growing body of data at $z\simeq 5-6$ but will also
include the emerging data at higher redshift to see how it agrees with
our model predictions.  A crucial ingredient in finding the model
parameters that best fit the observational data is the error budget on
that data. The commonly-used scatter due to Poisson fluctuations must
be supplemented by the spatial clustering of galaxy halos within the
volume of each survey. Here we will provide new error bars for
existing data and show that the cosmic variance due to clustering
dominates over Poisson fluctuations below a particular galaxy
luminosity for a given survey geometry. This understanding can be used
to optimize the flux sensitivity and strategies of future high-$z$
surveys.

The plan of the paper is as follows. In $\S$2 we introduce the basic
ingredients of our model for star forming galaxies at high-redshifts.
In \S 3 we calculate the error budget for observations of LBGs and
LAEs  at high redshift. The model is calibrated against existing data
for LBGs and LAEs at $z\simeq$5--6 in \S 4 and used to discuss the
emerging data at earlier epochs in \S 5. In $\S$6 we discuss the
implications  of our model for further observational campaigns with
current and  projected facilities.  In  \S 7, we summarize our
conclusions.
  
Throughout the paper, we have assumed a flat universe and
($\Omega_m,\Omega_\Lambda$)=(0.24,0.76) following results in
\cite{Spergel2006}.  All magnitudes are given in the AB system.

\section{A Physical Model for High Redshift Star Forming Galaxies }

The rationale of this paper is to empirically-calibrate, using the
data now available at $z\simeq 5-6$, the parameters of a simple model
that describes the evolving luminosity function of star-forming
galaxies (both LBGs and LAEs) over the quite short  time interval
corresponding to the redshift range 5$<z<$10. Such a  model can then
be used to make predictions for the upcoming  $z\simeq 7-10$ surveys.

The model assumes star formation at these early epochs is primarily
triggered by the well-understood rate at which dark matter halos
coalesce.  We assume that the ratio of baryons to total mass in halos
above some minimum mass \citep{WL06a} follows the universal value
$\Omega_b/\Omega_m$.  Baryons are subsequently converted to stars with
an efficiency given by $\rm f_\star$.  Following \cite{Loeb2005} and
\cite{WL06a}, we define the star formation timescale, $\rm t_{LT}$, as
the product of the star formation duty cycle, $\rm \epsilon_{DC}$, and
the cosmic time, $\rm t_{H}\equiv 2/3H$ at that redshift.  Using these
ingredients, the star formation rate $\rm {\dot{M}}_\star$ is related
to halo mass $\rm M_{halo}$ as follows
\begin{equation}
\rm {\dot{M}}_\star (M_{\rm halo}) = \frac{f_{\star}\times
  ({\Omega_b}/{\Omega_m})\times M_{halo}}{t_{LT}} .
\label{sfr_model}
\end{equation}

For comparison to LBG samples, the star formation rate defined above
is converted to the luminosity per unit frequency at 1500
\AA\ following the prescription presented in \cite{Madau98}: $\rm
L_{1500{\rm \AA}}=8.0\times 10^{27} ({\dot{M}_\star}/ {M_\odot~{\rm
    yr}^{-1}})~{\rm erg~s^{-1}Hz^{-1}}$.  This conversion factor
assumes a Salpeter initial mass function (IMF) of stars; if the IMF is
more top-heavy than the Salpeter IMF, the far-ultraviolet luminosity
will be greater for a given $\rm {\dot{M}}_\star$.

Comparison to LAE samples requires converting the star formation rates
derived above to a Ly$\alpha$ luminosity.  We do this assuming that
two-thirds of all recombining photons result in the emission of a
Ly$\alpha$ photon (case B recombination).  The ionizing photon
production rate is calculated from the star formation rate for a given
IMF and metallicity.  We fix the metallicity at 1/20 solar and assume
a Salpeter IMF which yields $\rm N_\gamma=4\times 10^{53}$ ionizing
photons per second per star formation rate in $\rm M_\odot/{\rm yr}$
\citep{Schaerer03}.  An  extreme top-heavy Population III IMF would
produce 3$\times$10$^{54}$ ionizing photons \citep{Bromm2001,
  Schaerer03}, and we will consider such an IMF in later sections.
Since the Ly$\alpha$ photons are assumed to be produced via
recombinations, only the fraction of ionizing photons which do not
escape into the intergalactic medium, ($\rm 1-f_{\rm ip}$), produce
Ly$\alpha$ photons.  Furthermore, only a fraction, $\rm
T_{{Ly}\alpha}$, of the emitted Ly$\alpha$ photons escape the galaxy
and are transmitted through the intergalactic medium (IGM).  With this
prescription the Ly$\alpha$ luminosity is related to the halo mass as
follows:
\begin{equation}
L_{{\rm Ly}\alpha} = \frac{2}{3} h\nu_{Ly\alpha} N_\gamma T_{{\rm
    Ly}\alpha}(1-f_{\rm ip}) {\dot{M}}_\star .
\label{lya_model}
\end{equation}
A substantial change in the IGM transmission parameter $T_{{\rm
    Ly}\alpha}$ is expected to signal the end of reionization.

We also consider a more advanced model, incorporating the effects of
supernova feedback on the luminosity function of star-forming
galaxies.  Since supernova feedback can significantly reduce the
efficiency of star formation in low-luminosity galaxies, it is
particularly important to consider when predicting the efficiency of
future surveys for high-redshift galaxies aimed at detecting
intrinsically fainter systems, rather than the luminous and rare
objects that have been detected thus far, Following the scaling
relations presented in \cite{Dekel03}, we assume there is a critical
halo mass at each redshift below which the star formation efficiency
begins to decrease due to feedback.  The star formation efficiency
(which we now call $\eta(M)$) is a function of halo mass,
\begin{equation}
\eta(M_{halo}) = 
\begin{cases}
\rm f_\star\left(\frac{M_{halo}}{M_{halo,crit}}\right)^{2/3} &
\text{M$_{\rm halo}<$M$_{\rm halo,crit}$}\\ f_\star  & \text{M$_{\rm
    halo}>$M$_{\rm halo,crit}$} \\  
\end{cases}
\end{equation}
where $\rm M_{halo,crit}$ represents the critical value.  At a given
redshift, the critical halo mass can be related to a critical halo
velocity.  In the local universe, observations suggest a critical halo
velocity of $\sim 100$ km s$^{-1}$ \citep{Dekel03}; we adopt this for
our high-redshift models, assuming that the physics of supernova
feedback depends only on the depth of the gravitational potential well
of the halos.

Finally, in \S 6 we also consider the expected evolution of  galaxy
sizes.  This is particularly important for gauging the efficiency  of
future surveys utilizing adaptive optics in detecting galaxies at
$z\gsim 7-20$.   If a galaxy is resolved in a particular observation,
the  signal-to-noise ratio for the detection increases as the  size of
galaxy decreases since a smaller aperture (with less  noise) is
required to encircle its flux.   Next-generation adaptive optics
systems on thirty-meter class ground-based  telescopes will offer a
resolution of $\gsim$9 milliarcseconds at 1.1$\mu$m,  corresponding to
$\simeq$50 pc at $z\simeq$7. With such exquisite angular resolution,
the size of early galaxies is likely to be a limiting factor in their
detection.

We assume that the extent of the stellar region of a galaxy at a
particular epoch is a constant fraction of the size of the host dark
matter halo.  The virial radius of a halo is given by
\begin{equation}
r_{vir}=\left[\frac{GM_{vir}}{100H^2(z)}\right]^{1/3}
\label{rvir}
\end{equation}
where M$_{vir}$ is the halo virial mass and $\rm H(z)=H_0[
  \Omega_{\Lambda,0} + (1-\Omega_{\Lambda,0}-\Omega_0)(1+z)^2
  +\Omega_0(1+z)^3]^{1/2}$ is the Hubble canstant at redshift $z$.  At
very high redshift, the halo virial radius scales as  $(1+z)^{-1}$ for
fixed halo mass.  Following the parameterization  presented in
\cite{Barkana2000}, a typical galaxy brighter than 1 nJy  between
$5\lsim z \lsim10$ will have a disk radius of 0\farcs1-0\farcs2, with
the range depending on the efficiency with which baryons are converted
to stars.  

Observations lend support to this simple scaling of  disk radius with
redshift: $z\simeq 2-6$ dropouts in the UDF  and Hubble Ultra Deep
Field Parallels (UDF-P)  are best fit by a $(1+z)^{-1}$ power law for
a  fixed luminosity  \cite{Bou04}.  The mean half-light radius of
0.3-1.0 $L_{\star,z=3}$  LBGs at $z\simeq 6$ is 0.8 kpc, corresponding
to 0\farcs1 at  $z\simeq 6$, consistent with the model presented in
\cite{Barkana2000}.   To make predictions for future observations in
\S 6, we thus assume  the mean size of galaxies of similar luminosity
scales as  0\farcs1$({1+z}/{7})^{-1}$.  

\section{The Effect of Variance in Deep Surveys}

Interpreting and planning observations of galaxies in the reionization
era  requires an accurate understanding of uncertainties arising from
both Poisson errors and fluctuations in the large-scale distribution
of  galaxies.  For example, one of the main motivations for galaxy
surveys at  $z\simeq $7--10 is the question of whether the imprint of
reionization  can be seen in the evolution of LAEs
\citep{Malhotra2006, Dijkstra2006} or  dwarf galaxies
\citep{BL06,Babich2006}. However, to answer these  questions, it is
important to ensure that the variance is less than the  claimed
evolution.  In this section, we develop the formalism necessary to
compute the variance for narrowband, dropout, and spectroscopic
lensing surveys, and we then apply this formalism to recent surveys.
An analysis of clustering variance was presented in
\cite{Somerville04}.  We improve upon two simplifications made in that
work, both of which we discuss below.

The variance due to Poisson errors is given by $\mathcal{N}_i$, where
$\mathcal{N}_i$ is the number of galaxies in luminosity bin $i$.  To
compute this we determine the number density of galaxies (either LBGs
or LAEs) as a function of luminosity using the best-fitting models
described in \S4, from which we can determine the predicted counts for
a given survey volume.  When characterizing the variance at redshifts
where data is sparse ($z\gsim 7$), we assume no evolution in the model
parameters from $z\simeq 6$.  This method clearly has its limitations
when probing new parameter space (e.g. low luminosities or high
redshifts) where the abundance of galaxies is not well-known.
However, given the lack of data, we consider it to be the simplest
approach.

The clustering of galaxies in overdense regions causes fluctuations in
galaxy counts, often referred to as cosmic variance.  Determining this
variance requires knowledge of the mass of the dark matter haloes that
host the observed galaxies.  If the correlation function of the galaxy
population is known the clustering variance can be predicted for a
given field of view.  Moreover, if the data set is sufficiently large,
the correlation function can be derived separately for bright and
faint galaxies, thereby showing how the dark matter halo mass (and
hence the clustering variance) varies with galaxy luminosity.  A
detailed spatial correlation function analysis is very challenging for
current observations at $z\gsim 6$; hence an alternative method is
needed for deriving the clustering variance.

Using the LBG and LAE model described in \S 2, the halo mass can be
determined as a function of galaxy luminosity.  The clustering
variance can then be calculated for a given halo mass by taking the
product of the variance of dark matter in the survey volume and the
bias factor associated with halos of a given mass $M$.  With this
method, we compute the clustering variance as a function of galaxy
luminosity for narrowband, dropout, and spectroscopic lensing surveys,
taking into consideration the survey geometry specific to each survey.  

First, we compute the variance of dark matter in a given  smoothing
window as follows: 
\begin{equation}
\sigma^2(r)= \int P(k) \tilde{W}^2({\bf k})\,d^3 {\bf k}
\label{variance}
\end{equation}
where $P(k)$ is the power spectrum of density fluctuations
extrapolated to $z$=0 as a function of wavenumber $k$ and
$\tilde{W}^2({\bf k})$ is the Fourier transform of the window function
in real space.  The form of window function depends on the survey
geometry; in the following subsections, we detail the window functions
adopted in our analysis.  Non-linear corrections to the power spectrum
and probability distribution become important if the variance is
larger than unity.  In computing the variance, we adopt the non-linear
power spectra using the halo model fitting functions presented in
\cite{Smith03}.  While fluctuations are in the linear regime, their
probability distribution is Gaussian.  However, in the non-linear
regime, the probability distribution function appears to be
well-described by a log-normal distribution \citep{Kayo01}.
Confidence intervals are thus determined via the geometric mean and
standard deviation; the one-sigma upper and lower bound are given by
$\exp(\mu+\sigma)$ and $\exp(\mu-\sigma)$, respectively, where $\mu$
and $\sigma$ are the mean and standard deviation of the logarithm of a
given density fluction, $\ln\,\delta$.  If the standard deviation of
the log-normal distribution is much less than one, $\sigma_{LN} \ll
1$, the probaility distribution function reduces to Gaussian with
confidence intervals given by $\mu\pm\sigma$.

The clustering variance in the distribution of {\it galaxies} is then
estimated by multiplying the variance in dark matter  by the halo
bias, which is defined as the ratio of the rms  fluctuations of haloes
to that of dark matter.  We adopt the halo bias formula  derived for
the ellipsoidal collapse model by \cite{Sheth01}.

A key assumption in the clustering variance formalism described above
is that there is not more than one galaxy per halo.  More complex
occupation numbers are possible, but since cooling is efficient at
such high redshifts, the simplest case is that with one galaxy per
halo.

Now we introduce two key differences between our approach and that of
\cite{Somerville04}.  First, the survey geometry was assumed to be
spherical in \cite{Somerville04}.  However, different survey
geometries may have substantially different power spectra
\citep{Kaiser91}.  This is a particular concern for strong lensing
surveys that utilize longslit spectroscopy \citep{Santos04,Stark07b}.
Second, the observed number density of the population was assumed to
be equivalent to the number density of the underlying dark matter
halos \citep{Somerville04}.  This assumption could be in error if the
star formation duty cycle, $\epsilon_{\rm DC}$, is significantly
smaller than unity.  In this case, the observed number density of
galaxies would be less than that of their host dark matter haloes by a
factor of $\epsilon_{\rm DC}$.  Depending on the slope of the mass
function, this would overestimate or underestimate the cosmic
variance.

\begin{figure}
\figurenum{1} \epsscale{.95}
\plotone{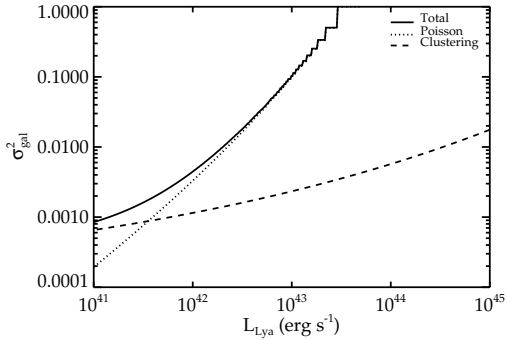}
\caption{Variance in a narrowband Ly$\alpha$ survey.  The total
  variance (solid line) is the sum of the variance from Poisson noise
  (dotted line) and clustering fluctuations (dashed line).  The survey
  specifications adopted in the left panel are equivalent to Subaru
  survey for Ly$\alpha$ emitters at $z=5.7$ \citep{Shimasaku06}.  The
  clustering fluctuations in such a narrowband survey are very small
  (less than 6\%) in the luminosity range over which Lya emitters are
  detectable.  }
\label{plot_sigma_nb}
\end{figure}

In the following subsections, we apply the formalism described above
to recent surveys for LBGs and LAEs.  For each survey, we compute the
typical  amplitude of clustering and Poisson fluctuations in the
relevant observational regime.  In addition, we determine the
cross-over  luminosity, L$_c$, defined as the luminosity above which
the Poisson error  dominates that from clustering fluctuations.

\subsection{Variance in Narrowband Surveys}

The geometry of a narrowband LAE can be approximated as a rectangular
parallelepiped with a comoving volume of $a_x a_y a_z$, where $a_x$
and $a_y$ are the comoving distances corresponding to the areal field
of view of the survey and $a_z$ is the comoving line-of-sight distance
of the survey.  The window function in k-space is simply the Fourier
transform of a rectangular top-hat with dimensions corresponding to
the survey geometry:
\begin{equation}
\tilde{W}(k_x,k_y,k_z) = \frac{\sin(k_x a_x/2)}{(k_x
  a_x/2)}\frac{\sin(k_ya_y/2)}{(k_y a_y/2)}\frac{\sin(k_za_z/2)}{(k_z
  a_z/2)} .
\label{window_nb}
\end{equation}

Using the window function defined above, we evaluate the expected
variance in narrowband surveys for LAEs at $z\simeq 6$.  We focus our
analysis on the narrowband surveys conducted at $z=5.7$ and $z=6.5$ in
the Subaru Deep Field (SDF).  Using the wide-format, Suprime-Cam
\citep{Miyazaki02} on the Subaru Telescope, the observations sample an
area of 34\arcmin$\times$27\arcmin\ in the NB816 and NB921 narrowband
filters.  The central wavelength and FWHM of the NB816 and NB921
filters are (8150~\AA, 120~\AA) and (9196~\AA, 132~\AA) respectively.

In Figure \ref{plot_sigma_nb}, we plot the variance in the $z=5.7$
narrowband survey of the SDF.  The Ly$\alpha$ luminosities probed in
the Subaru survey range from $\simeq\rm10^{42}-10^{43}~{\rm
  erg~s^{-1}}$. The clustering variance is less than $\simeq$0.01 over
this luminosity range, resulting in less than 10\% uncertainty in the
observed counts.  Poisson errors dominate the clustering errors for
sources brighter than the cross-over luminosity of 10$^{41.7}$ erg
s$^{-1}$ (Table 1).  Since this is intrinsically fainter than  the
luminosity limit of the SDF, this survey is dominated by Poisson
errors. 

\subsection{Variance in Lyman-break Surveys}

We focus our dropout survey analysis on the two
16\arcmin$\times$10\arcmin~ Great Observatories Origins Deep Surveys
(GOODS) of the Hubble Deep  Field North (HDF-N) and Chandra Deep Field
South (CDF-S) and the  3.4\arcmin$\times$3.4\arcmin~ Hubble Ultra Deep
Field (UDF).  The  redshift distribution of dropouts depends on the
filters and color-cuts  used in their selection.  While the typical
color-selection criteria  for $i-drops$ select galaxies between
$z=5.5$ and $z=7.0$ \citep{Bunker04},  the effective distance sampled
along the line-of-sight is less than  the total comoving radial
distance in this redshift interval because  of incompleteness arising
from objects being scattered faintward of  the magnitude limit or out
of the color-selection window.  This  incompleteness has been
quantified in both GOODS and the UDF  \citep{Bunker04,Bou06b},
allowing an effective volume to  be derived for each survey.  We
approximate the   geometry of the LBG survey as a rectangular
parallelepiped with  a characteristic line-of-sight distance equal to
the ratio of the  effective volume and the survey area and with
dimensions in the  plane of sky corresponding to the field-of-view of
the survey.   

\begin{figure}
\figurenum{2} \epsscale{1.1} \plotone{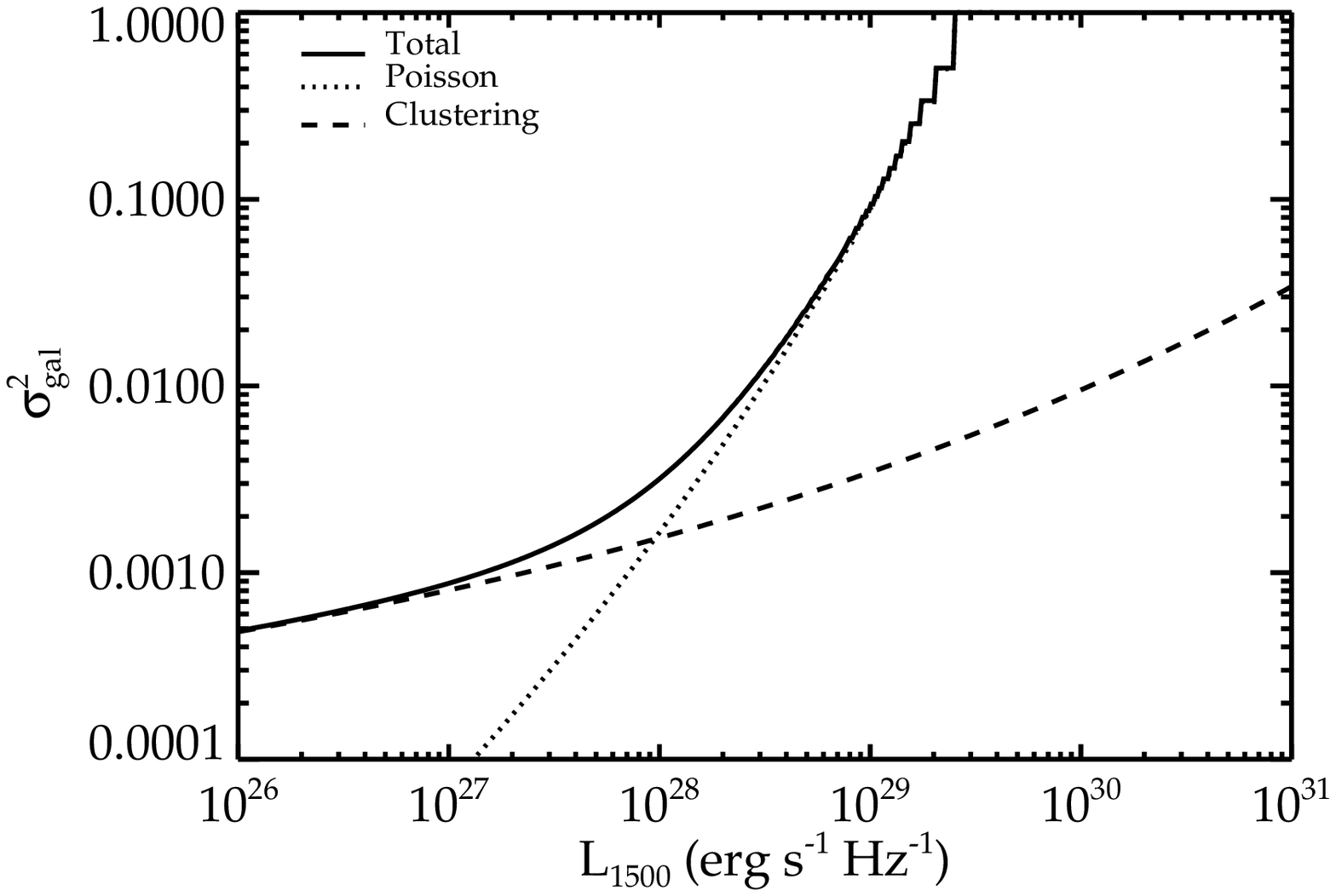} \plotone{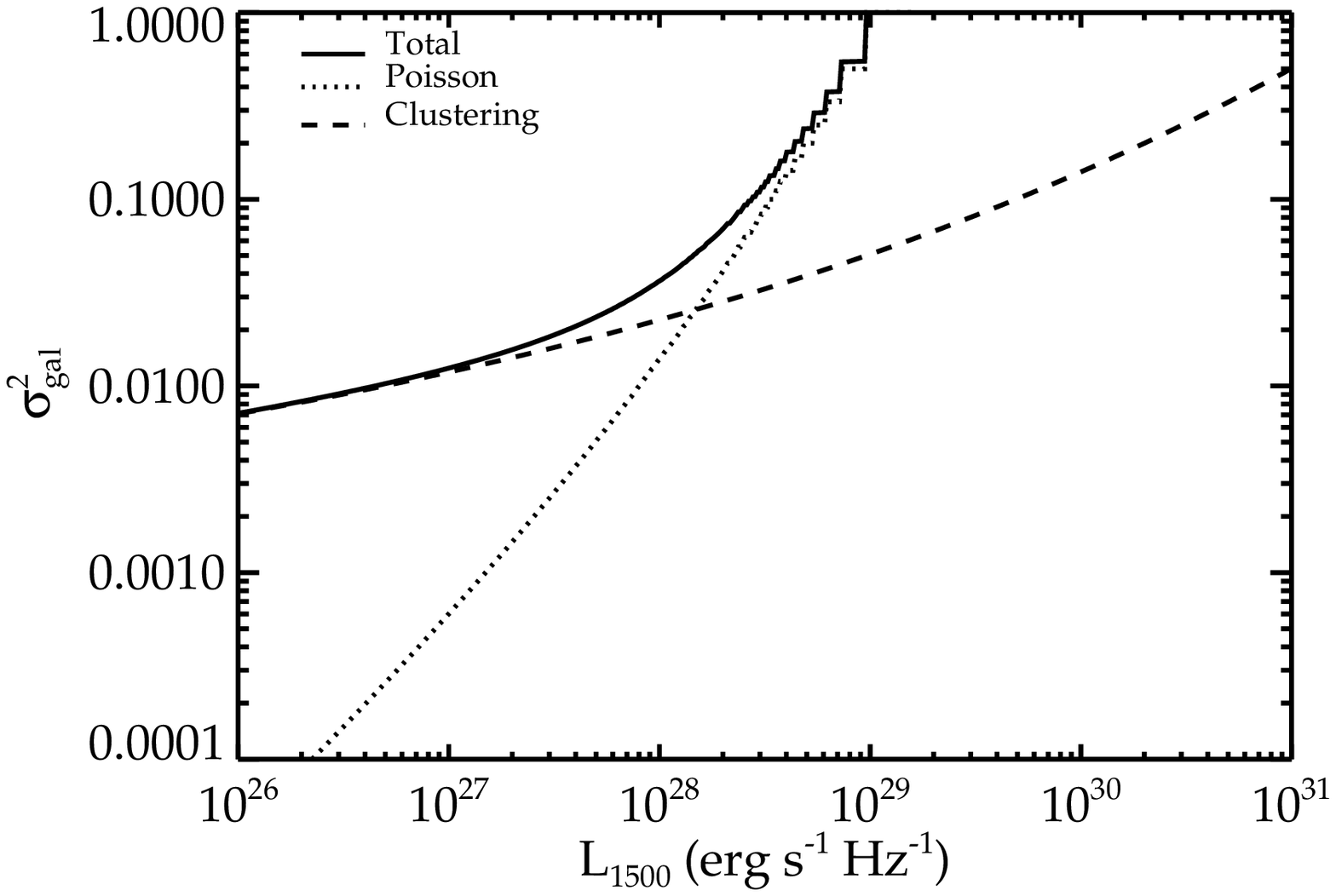}
\caption{Variance in dropout LBG survey. The total  variance (solid
  line) is the sum of the variance from Poisson noise  (dotted line)
  and clustering fluctuations (dashed line).  In  the case of a survey
  such as the HST-GOODS observations of the CDF-S and HDF-N (limiting
  luminosity of $\rm 5\times10^{28} erg s^{-1} Hz^{-1}$),  the
  clustering fluctuations of $z\simeq 6$ LBGs are greater than
  $\simeq$6\% (top panel). In the Hubble Ultra Deep field (limiting
  luminosity of $\rm 8\times10^{27} erg s^{-1} Hz^{-1}$), clustering
  fluctuations result in a slightly higher uncertainty of
  $\gsim$15-20\% (bottom panel).  }
\label{plot_sigma_drop}
\end{figure}

The variance of the GOODS and UDF surveys for $z\simeq 6$ LBGs is
presented in Figure \ref{plot_sigma_drop}. The GOODS survey is
sensitive to sources brighter than 5$\times$10$^{28}$ erg s$^{-1}$
Hz$^{-1}$ (corresponding to a 10$\sigma$ limit of $z'_{AB}$=26.5).  At
this limit, the clustering variance is only 0.003, and the total
variance is dominated  by the Poisson term.  The clustering
fluctuations are slightly lower than those estimated in
\cite{Somerville04}, due largely to the more  realistic source
geometry.  In the UDF, both the clustering and Poisson  error are
larger due to the smaller survey volume.  The 8$\sigma$ limiting far
UV luminosity in the UDF is 8$\times$10$^{27}$ erg s$^{-1}$ Hz$^{-1}$
\citep{Bu04}.  The total variance near this limit is dominated by
clustering fluctuations, which contribute an uncertainty of
$\sim$15-20\% to the observed densities.  

The field-to-field fluctuations of $z\simeq 6$ LBGs have been measured
in \cite{Bou06} by degrading the depth of the UDF to that of the  two
UDF-parallel fields and subsequently comparing the number of
i-dropouts in each field.  The density of i-dropouts selected in the
(degraded) UDF is similar to that in the first parallel (50.2 and
42.6, respectively) but is significantly greater than that in the
second parallel (27.8 vs. 11.4).  The latter comparison implies
field-to-field variations on 7 arcmin$^2$ at the magnitude limit
($z_{850} = 28.6~\rm{at}~8\sigma$) of the second UDF parallel are
$\simeq 40$\%.  The observed field-to-field variations result from
both Poisson (19\% for the degraded UDF and 30\% for the second UDF
parallel) and clustering fluctuations (19-26\% over 7 arcmin$^2$).
When these uncertainties are accounted for, the two measurements  are
consistent at the 1.3$\sigma$ level. 

\subsection{Variance in Lensed Longslit Spectroscopic Surveys}

The geometry of a longslit spectroscopic survey can be approximated as
a rectangular parallelepiped.  However, in the case of a strong
lensing longslit spectroscopic survey, the geometry is potentially
slightly more complex.  For a lensing survey, the slit geometry only
corresponds to the {\it image} plane; however, we are interested in
the geometry of the survey in the {\it source} plane, which can be
calculated accurately via ray tracing if a reliable mass model is
available.  While the source plane geometry depends on the location of
the longslit relative to the critical line, for typical  clusters
studied in \cite{Stark07b} it is well-approximated by a
rectangular-parallelepiped (J. Richard 2006, private
communication). The source plane area is reduced by a factor of the
lensing magnification, $\mathcal{M}$; further, the magnification is
not isotropic and is strongest perpendicular to the cluster critical
line.  Hence, assuming the longslit is oriented along the cluster
critical line, the slit-width is compressed more than the slit-length.
For the computations that follow, we assume the source plane slit
width, $a_x$, is related to the image plane slit width, $a'_x$ by
$a_x$=${a'_x}/{\sqrt{2\mathcal{M}}}$, and likewise, the source plane
slit width is given by $a_y$=${a'_y}\sqrt{2/{\mathcal{M}}}$, in
agreement with typical slit positions from \cite{Stark07b}.

As with the narrowband survey, the appropriate window function is a
three-dimentional rectangular top-hat in real space, which in k-space
corresponds to the product of three sinc functions (see Equation
\ref{window_nb}).

A typical near-infrared spectrometer has dimensions of  0.76\arcsec
$\times$ 42\arcsec \citep{Stark07b},  which at $z\simeq 9$ corresponds
to a comoving distance of  6 kpc$\times$610 kpc, assuming a median
magnification factor of  $\mathcal{M}$=20.  In each cluster, an area
is mapped out around the critical line; assuming six slit positions
are observed, this results in a total survey area of 0.02 Mpc$^2$ per
cluster.    If observations are conducted in the J-band between
$z=8.5$ and  $z=10.4$, the comoving line-of-sight distance spanned is
479 Mpc.  We compute the clustering variance expected over the total
survey volume for a fifteen cluster survey.  In the near future,
significantly more clusters will become available for strong lensing
surveys \citep{Ebeling03}.

\begin{figure}
\figurenum{3} \epsscale{.95} \plotone{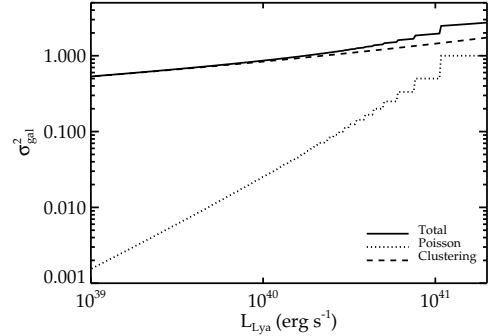}
\caption{Variance in a $z=10$ lensed longslit spectroscopic survey.
  The total variance (solid line) is the sum of the variance from
  Poisson noise  (dotted line) and clustering fluctuations (dashed
  line).  The  uncertainty from clustering fluctuations is
  significantly greater than the Poisson noise for all luminosities
  and is nearly 100\% for sources  with Ly$\alpha$ luminosities of
  10$^{40}$-10$^{41}$ and erg s$^{-1}$.  }
\label{plot_sigma_slit}
\end{figure}

The clustering fluctuations are significantly larger in the
spectroscopic lensing survey than in either the traditional narrowband
or dropout surveys due to the  much smaller survey volumes.  While
such surveys may offer the only prospect of detecting galaxies at
$z\simeq 10$, clearly there will still be large uncertainties in their
abundance due to cosmic variance.  The development of larger and more
sensitive near-infrared spectrometers is necessary to increase the
survey volume obtainable in a reasonable time allocation.

\section{Model Calibration Using $z\simeq5-6$ Observations}

We are now in a position to use our model and our improved
understanding of the effects of cosmic variance to constrain its
parameters using observations of the luminosity functions at $z\simeq
5-6$. By doing this independently for LBGs and LAEs we will hopefully
gain valuable insight into the physical differences between these two
star-forming populations.

\subsection{Lyman-Break Galaxies}

Lyman-break galaxies are perhaps more straightforward to model than
Ly$\alpha$ emitters because of the complex resonant interaction of
Ly$\alpha$ photons with neutral hydrogen which occurs in the latter
population.  In our model of LBGs described in \S2, there are two
free-parameters:  the star formation efficiency $f_\star$ and the duty
cycle  $\epsilon_{\rm DC}$.  First we determine these parameters by
reproducing  the observations at $z\simeq 6$. We later use this model
to consider whether the emerging data at  $z\simeq 7-10$ requires any
adjustment.  Significant evolution in the model parameters between the
two redshifts might signify some external phenomenon, such as the
reionization of the intergalactic neutral hydrogen. Alternatively it
could cast doubt on the reliability of the observations.

We compute a grid of LBG luminosity functions by varying $f_\star$ and
$\epsilon_{\rm DC}$.  The comoving number density of galaxies
predicted by the models, $n_{\rm mod}$, is compared to the observed
value, $n_{\rm obs}$, in each of the N luminosity bins, and a
likelihood of a given set of parameters is defined such that
$\rm\mathcal{L}(f_{\star},\epsilon_{\rm DC})=\exp[-0.5\chi^2]$, where
$\rm\chi^2=\Sigma_{i=1}^N (n_{{\rm obs},i}-n_{{\rm mod},i}$).  The
1-sigma  uncertainty in the observed densities include the
contribution from  cosmic variance (see \S3) in addition to that from
Poisson noise. 

\begin{figure}
\figurenum{4} \epsscale{1.1} \plotone{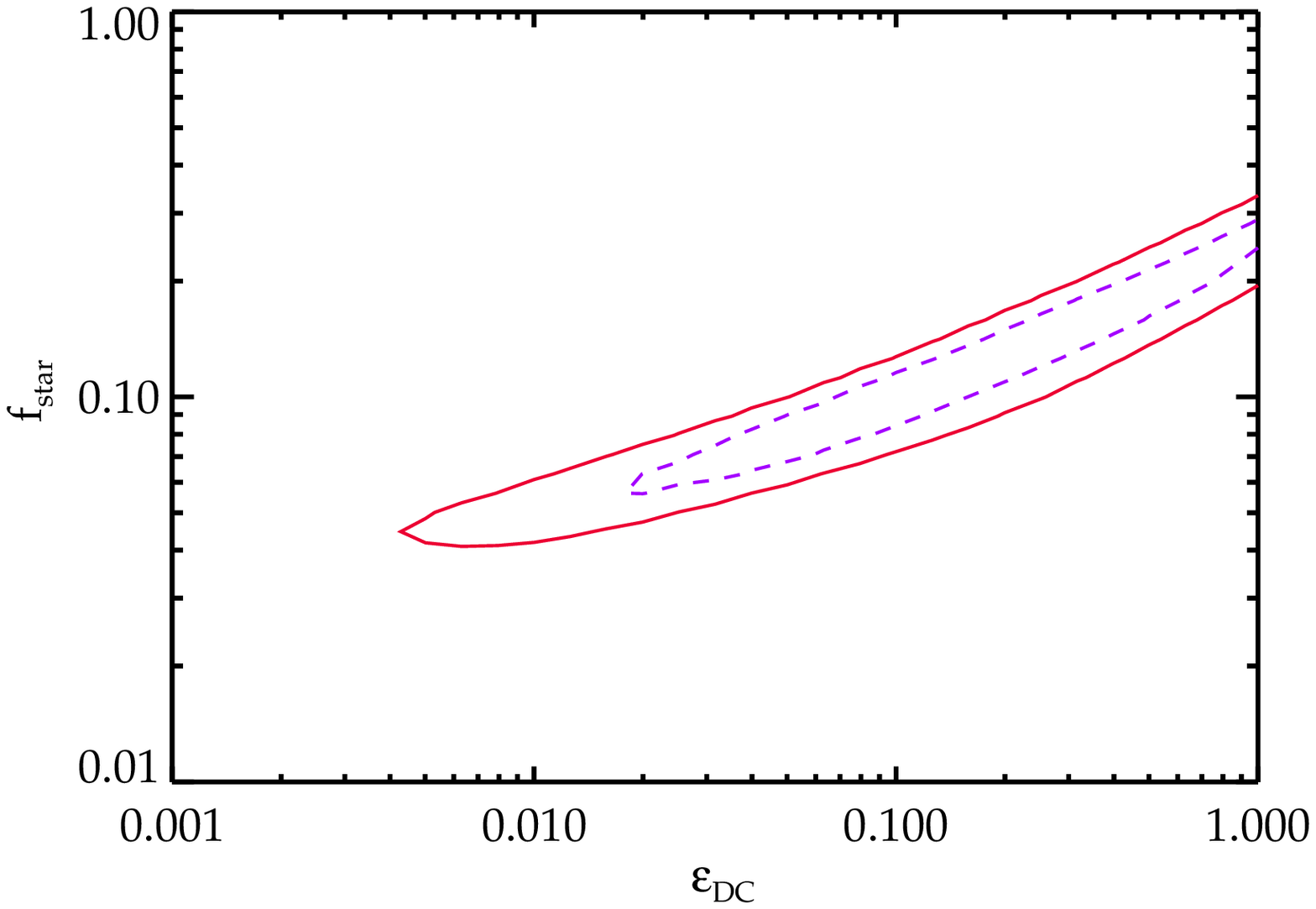} \plotone{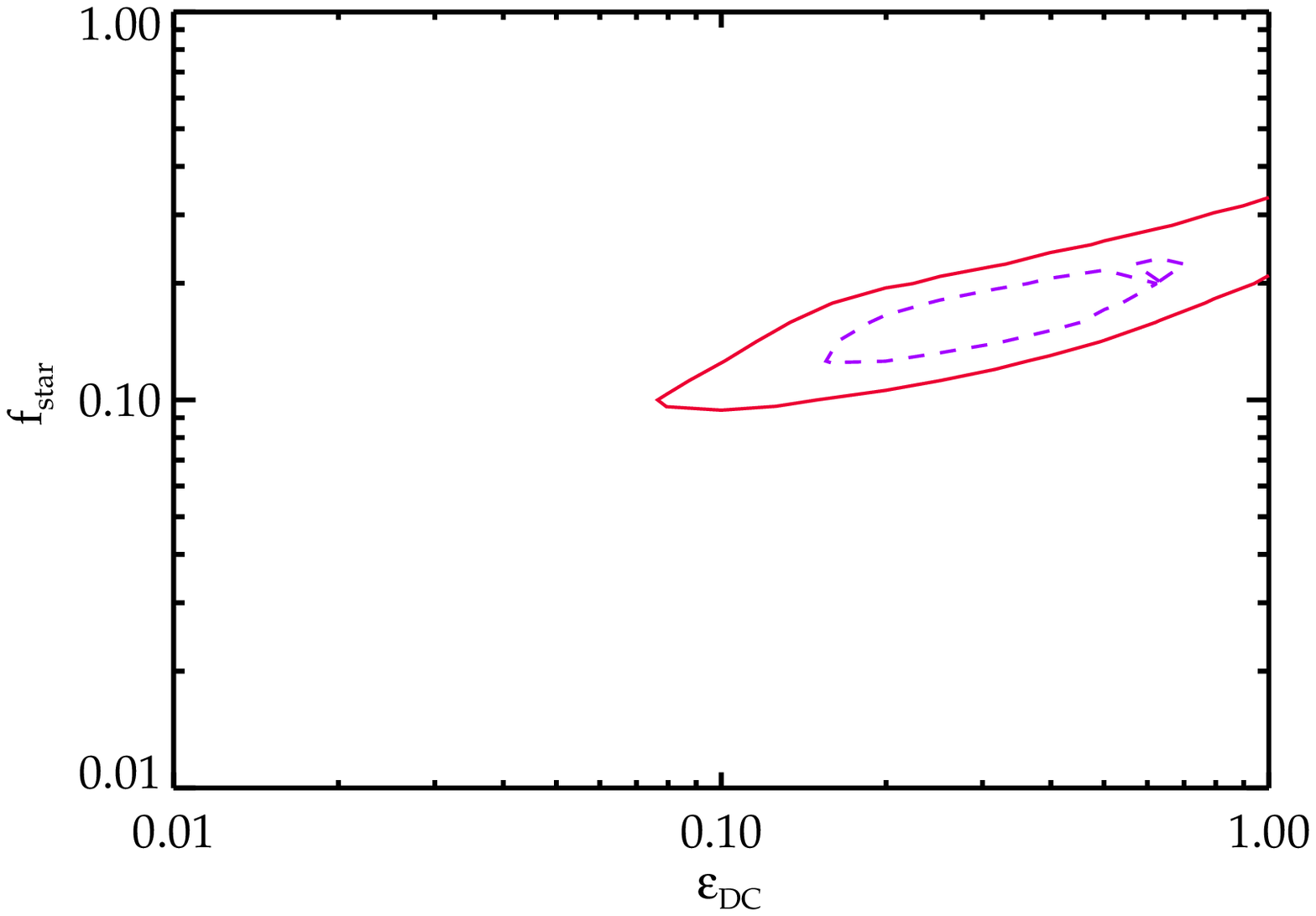}
\caption{{\it Top:} Confidence intervals on star formation efficiency
  $f_\star$ and duty cycle $\epsilon_{\rm DC}$ in a simple theoretical
  model for the observed abundances of LBGs at z=6.  The likelihood
  contours are 64\% (blue dashed line) and 26\% (red solid line) of
  the peak likelihood.  {\it Bottom:} Same as above  except with
  supernova feedback included in model.}
\label{plot_lbg_model}
\end{figure}

We first apply our model to the observed abundance of LBGs at $z\simeq
6$, as compiled by \cite{Bou06}.  In Figure \ref{plot_lbg_model}, we
show the likelihood contours at 64\% and 26\% of the peak likelihood
(corresponding to 1- and 2-$\sigma$ for a Gaussian distribution).  The
maximum likelihood and 1-$\sigma$ confidence intervals are
$(f_{\star},\epsilon_{\rm
  DC})=[0.13^{+0.15}_{-0.07},0.20_{-0.18}^{+0.80}]$, in reasonable
agreement with a similar fit to these observations in \cite{WL06a}.
When supernova feedback is allowed to decrease the star formation
efficiency in low-mass halos (see \S2 for details), the best-fit
parameters change slightly: $(f_{\star},\epsilon_{\rm
  DC})=[0.16^{+0.06}_{-0.03},0.25_{-0.09}^{+0.38}]$ and the likelihood
increases by almost a factor of two.  The strong degeneracy between
the  duty cycle and the star formation efficiency arises because an
increase in the star formation efficiency requires a longer star
formation timescale (and hence larger star formation duty cycle) to
produce the same far-UV luminosity for a given halo mass.

Although there is some degeneracy between the best-fitting star
formation efficiency and duty cycle at $z\simeq 6$, the range of
values can be physically understood.  A duty cycle of 20\% at $z\simeq
6$ corresponds to a star formation lifetime of $\simeq 200$ Myr which
is  only slightly larger than the dynamical time of virialized halos
(or the duration of equal-mass mergers) at that redshift.  Our simple
model thus suggests that at $z\simeq 6$, star formation is proceeding
on roughly the same timescale it takes virialized baryons to settle to
the center of the galaxy.  A star formation efficiency of $\sim 13\%$
is reassuringly similar to the ratio between the global mass density
in stars and baryons in the present-day Universe \citep{Fukugita98}.

An independent check on the inferred duty cycle could conceivably be
obtained if the spectral energy distribution was known for a large
sample of $z\simeq 5-6$ LBGs.  Fitting these with a grid of population
synthesis models \cite{BC03} allows, in principle, the estimation of
the stellar mass, dust extinction, and luminosity-weighted age of
representative galaxies. Unfortunately, the ages inferred via this
technique have large systematic uncertainties due to the inability of
the population synthesis models to constrain the past star formation
history \citep{Eyles2006, Shapley05}.  Taking the star formation
histories that minimize the $\chi^2$ fit to the SEDs of LBGs at
$z\simeq 6$, \cite{Eyles2006} find a median age of 500 Myr for those
objects detected in the rest-frame optical with Spitzer (and hence the
most massive objects).  A stacking analysis of the least massive LBGs
in their survey indicates that these objects have ages of $\simeq $60
Myr.  Recalling that the duty cycle is equal to the star formation
timescale divided by the cosmic time, these inferences suggest that
the duty cycle lies in the range 6-50\% and perhaps increases with the
mass of the galaxy.  

A further check is provided by limited data on the clustering of LBGs.
The halo masses probed in the \cite{Bou06} compilation in the GOODS
and UDF surveys are $7\times10^9~\rm{M_\odot}$--$3\times 10^{11}~
\rm{M_\odot}$  according to the simple model we have adopted.  These
values are  consistent with clustering analysis of $z\simeq 6$ LBGs in
GOODS, which  suggest that the hosting dark matter haloes are $\sim
10^{11}~\rm{M_\odot}$ \citep{Overzier06}.  Reionization would be
accompanied by a dramatic increase  in the cosmological Jeans mass and
and therefore in the minimum galaxy mass \citep{WL06a}.  A direct
detection of this effect requires finding galaxies in dark matter
haloes over an order of magnitude less massive than those probed in
current surveys \citep{BL06}.  

\begin{figure}
\figurenum{5} \epsscale{0.95} \plotone{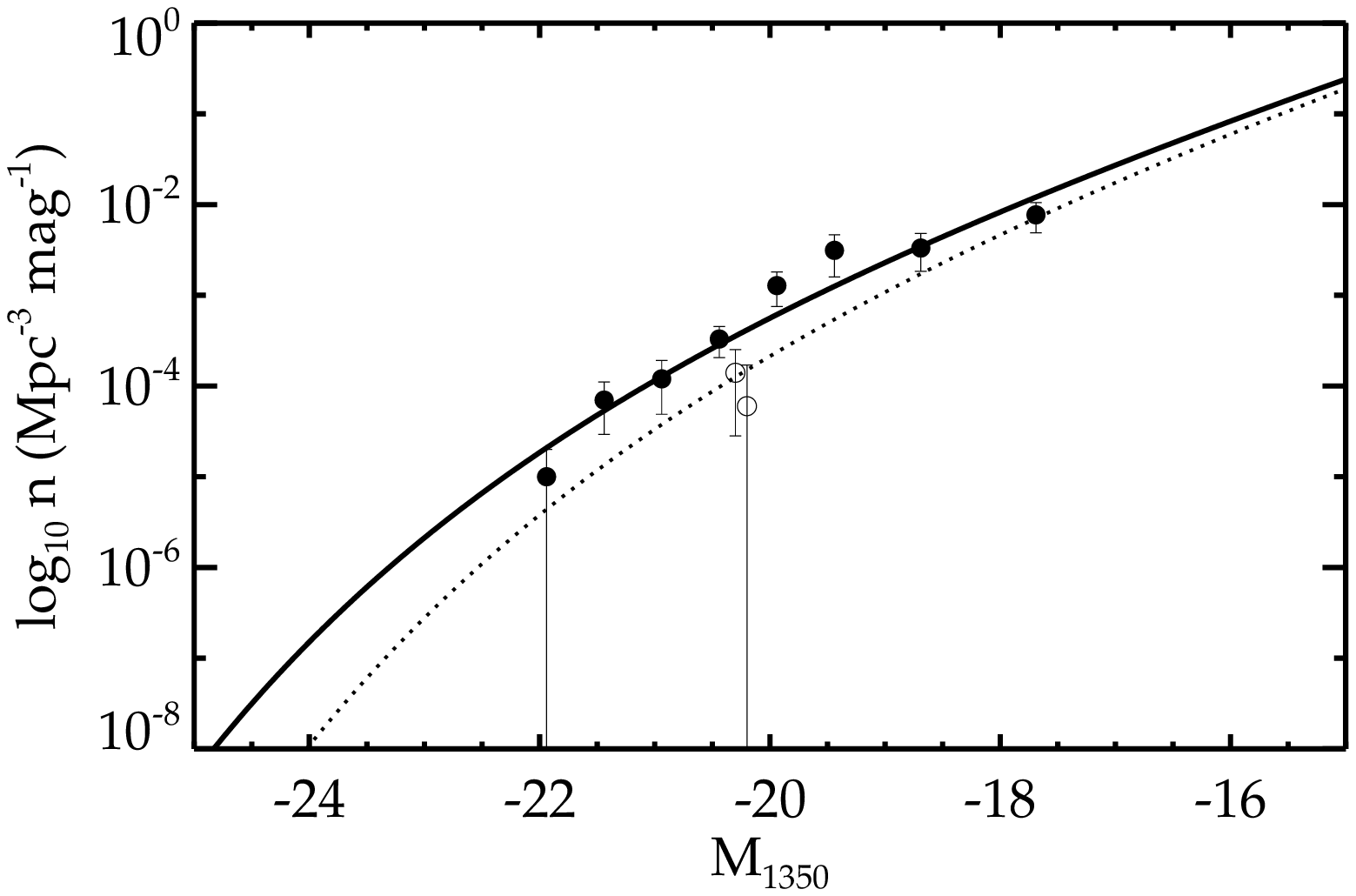}
\caption{The LBG luminosity function at z=6 (solid line) and z=7.6
  (dotted line) obtained using the model parameters that maximize the
  likelihood at z=6.  The solid and open circles correspond to
  observed LBG abundances at z=6 \citep{Bou06} and z=7.6
  \citep{Bou06b}. The two datapoints at z=7.6 (offset horizontally for
  clarity) correspond to more  and less conservative selections of
  z-drops in \cite{Bou06b}.  While there is not yet much data at
  z=7.6, the existing data at these two redshifts can be fit without
  any evolution in the fit parameters $f_\star$ and $\epsilon_{\rm
    DC}$.  }
\label{plot_lbg_lf}
\end{figure}

One important implication of our best-fit duty cycle is that
$\simeq$80\%  of dark matter halos of a given mass are not traced by
LBGs.  The ``missing''  dark matter halos may have gone through bursts
of star formation at  earlier times and may be currently quiescent.
However, this does  not mean that 80\% of the stellar mass is missing
from observations  at $z\simeq 6$.  Rather, the gas in the ``missing''
dark matter halos  may not have cooled sufficiently to be forming
stars rapidly  enough to be selected as LBGs and thus may not be
significant repositories  of stellar mass.  The first option suggests
that there may  have been a significant amount of star formation at
earlier times.  This is evidenced by observations of LBGs at $z\simeq
6$ with stellar  masses as great as a few $\times 10^{10}
\rm{M_\odot}$ and ages of 200-700 Myr \citep{Eyles05,Eyles2006}.
Given the current observed star formation rate of these galaxies, the
past star formation rate had to have been higher at earlier times.
Taken together, these observations and the 20\% duty cycle inferred
from the luminosity function of the LBGs suggest that the purported
deficity  of ionizing photons compared to taht required for
reionization  \citep{Bu04,Bou06} at $z\simeq 6$ could be accounted for
by earlier star  formation.

\subsection{Ly$\alpha$ emitters}

We now use a procedure similar to that described for the LBGs to model
the LAEs.  The key difference is that we must also consider the
fraction of Ly$\alpha$ photons that escape from the galaxy and IGM,
$T_\alpha$.  We generate a grid of models at $z=5.7$ and $z=6.5$ with
the duty cycle, $\epsilon_{\rm DC}$ ranging from 10$^{-3}$ and 1 and
the  product of the star formation efficiency and Ly$\alpha$ escape
fraction, $\rm f_\star T_{\alpha}$ spanning between 10$^{-3}$ and 1.
Each model is  compared  to the observed abundances, and the
likelihood is then determined for each  model in an identical fashion
to that discussed for the LBGs.  We first  perform this procedure for
our simple model and then examine it in the context of a model
including supernova feedback.

\begin{figure}
\figurenum{6} \epsscale{1.1}  \plotone{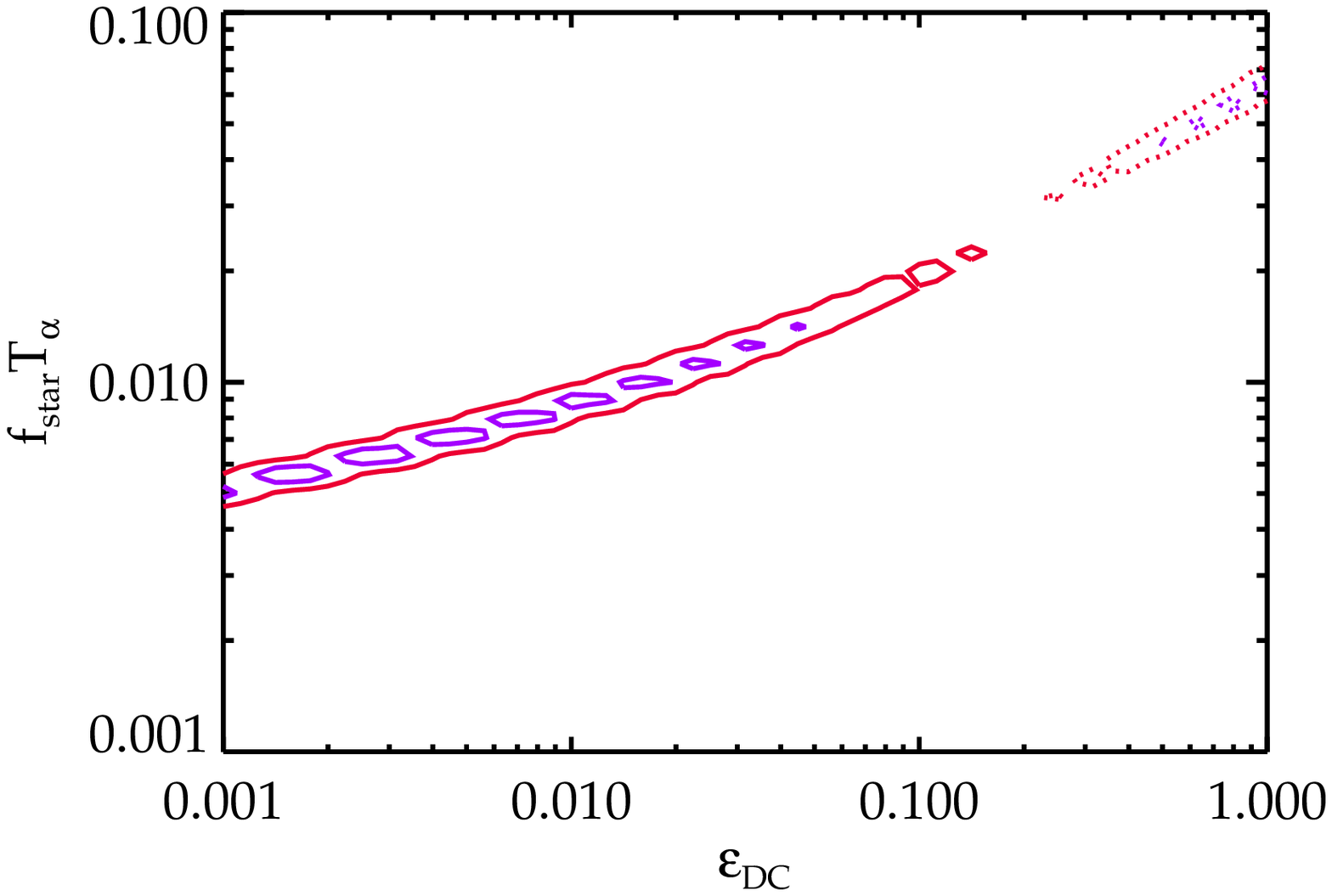} \plotone{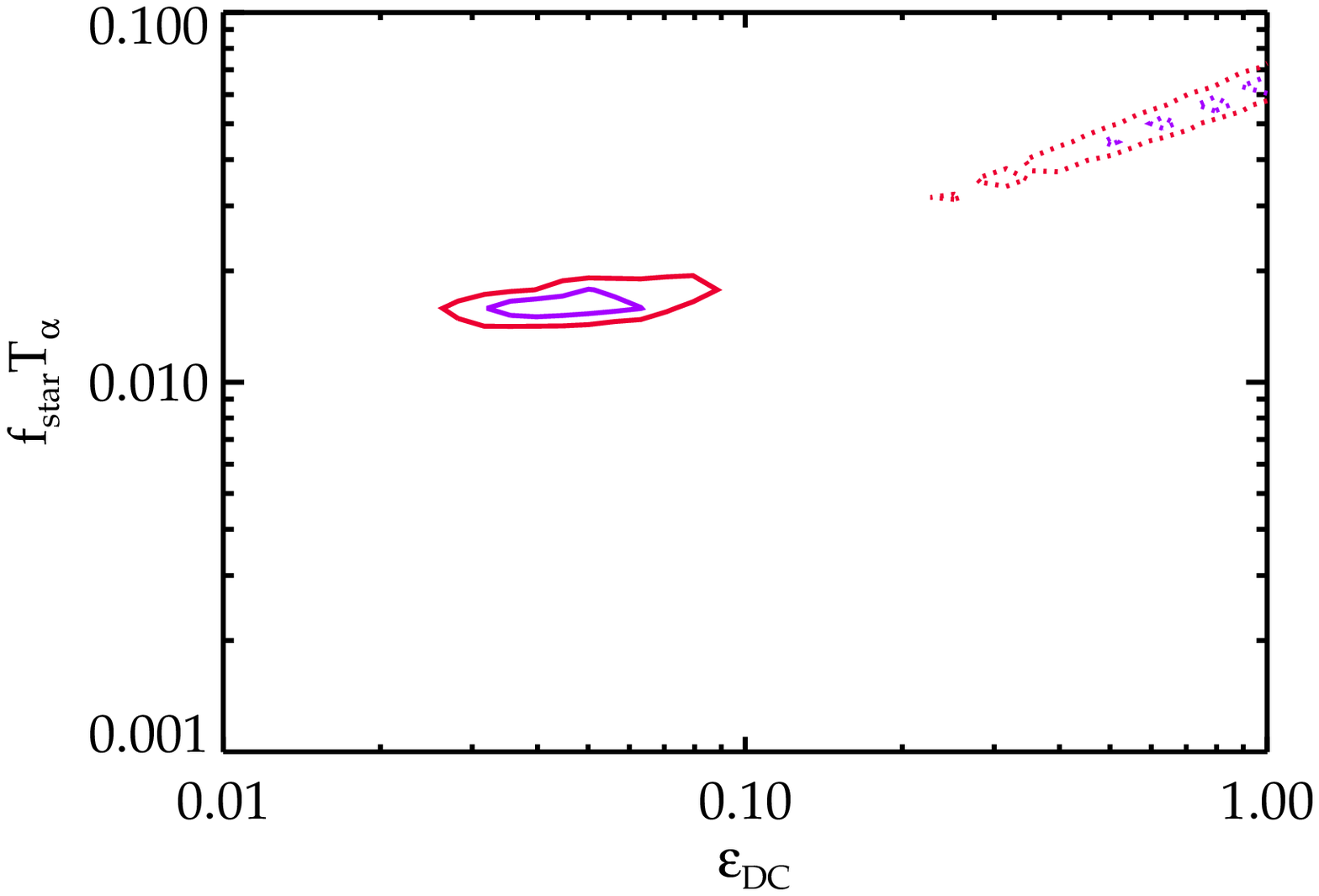}
\caption{{\it Top:} Confidence intervals on free parameters in
  analytic fit of observed LAE abundances at z=5.7 (solid lines) and
  z=6.5 (dotted lines).  The likelihood contours are 64\% (blue) and
  26\% (red) of the peak likelihood. The parameters that maximize the
  likelihood are ($ \epsilon_{DC},
  f_{star}T_{\alpha}$)=(0.0016,0.0056) at z=5.7 and ($ \epsilon_{DC},
  f_{star}T_{\alpha}$)=(1.0,0.063) at z=6.5.  {\it Bottom:}  Same as
  Figure \ref{plot_lya_confidence}a with the addition  of a simple
  prescription for supernova feedback.}
\label{plot_lya_confidence}
\end{figure}

Likelihood contours are presented in the top panel of Figure
\ref{plot_lya_confidence} for $z=5.7$ (solid contours) and $z=6.6$
(dotted contours).  As with the fit to the LBGs, there exists a strong
degeneracy between $\epsilon_{\rm DC}$ and $f_\star T_\alpha$.  The
best-fitting parameters (with associated one-sigma uncertainties) are
($\rm \epsilon_{DC},f_{\star}T_{\alpha})=(0.0016^{+0.0431}_{-0.0006},
0.0056^{+0.0085}_{-0.0006})$ at $z=5.7$ and ($\rm
\epsilon_{DC},f_{\star}T_{\alpha})=
(1.0^{+0.0}_{-0.5},0.063^{+0.004}_{-0.018})$ at $z=6.5$.  The $z=5.7$
data are significantly better fit (factor of 4 greater maximum
likelihood at $z=5.7$) by the advanced model including supernova
feedback (bottom panel of Figure \ref{plot_lya_confidence}).  The
best-fitting parameters at $z=6.5$ remain unchanged, while those at
$z=5.7$ change slightly:  ($\rm
\epsilon_{DC},f_{\star}T_{\alpha})=(0.040^{+0.023}_{-0.004},
0.016^{+0.0019}_{-0.0017})$.  Since the model with supernova feedback
provides a better fit to the $z=5.7$ data, we focus our discussion on
the model parameters derived in this fit, rather than the most simple
model, in our discussion below.

\begin{figure}
\figurenum{7} \epsscale{.95}  \plotone{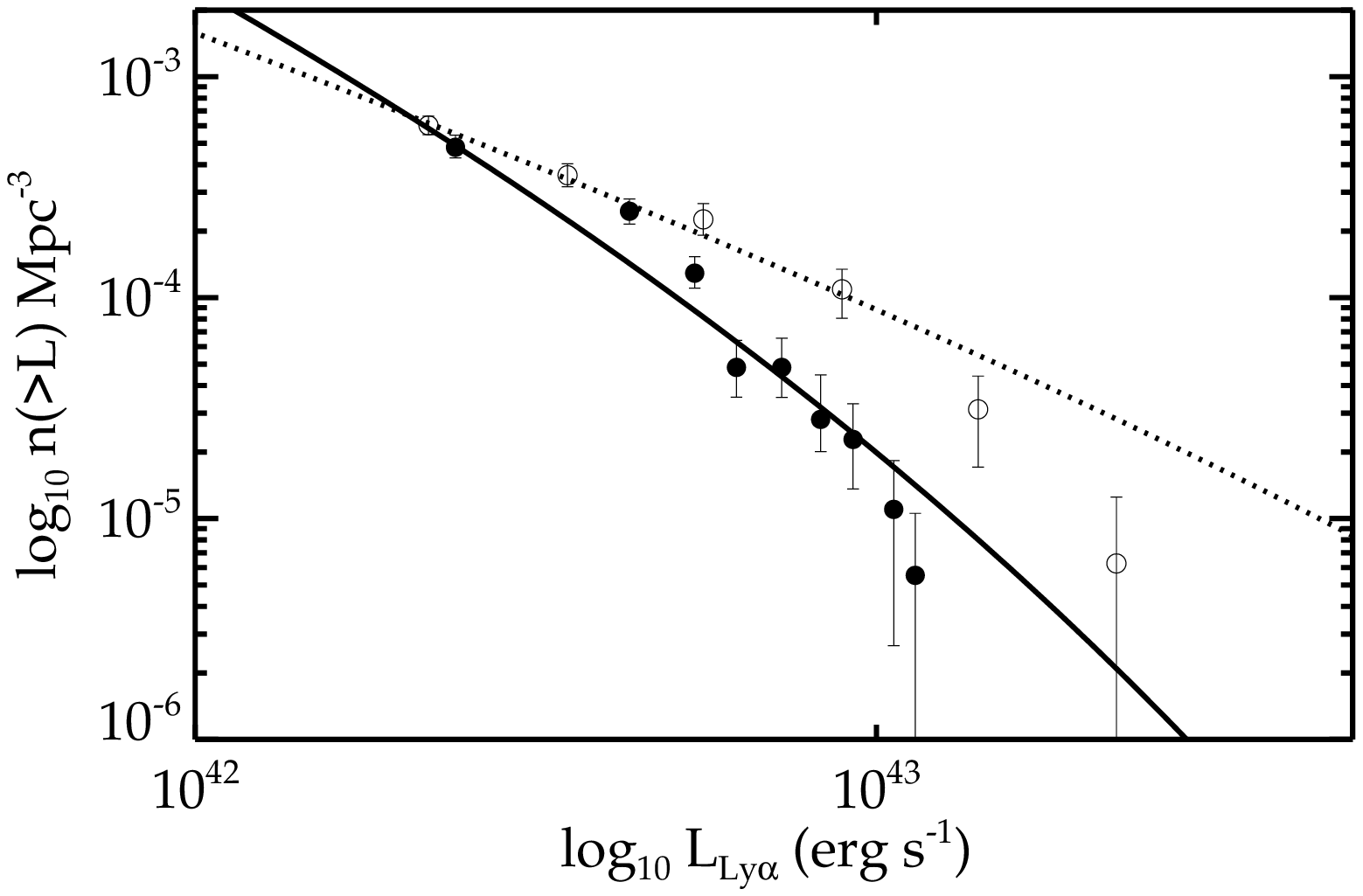} \plotone{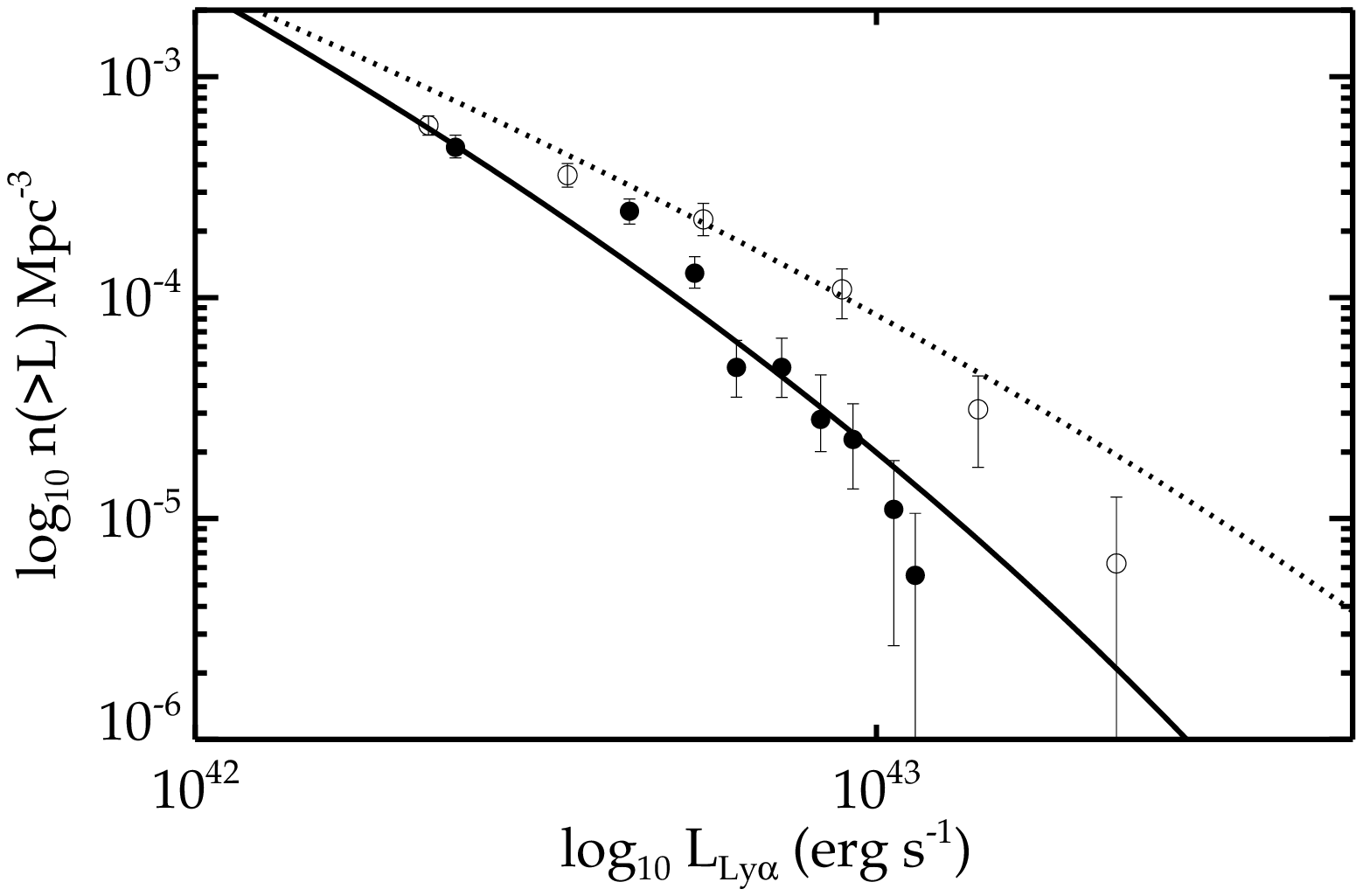}
\caption{{\it Top:} The Ly$\alpha$ luminosity function at z=5.7
  (dotted line) and z=6.5 (solid line) obtained using the model
  parameters that maximize the likelihood.  Open and solid circles are
  observed abundances at $z=5.7$ from \cite{Shimasaku06} and at
  $z=6.5$ from \cite{Kashikawa06}. {\it Bottom:} Same as above but for
  best-fitting parameters to model that now includes a simple
  prescription for supernova feedback.  }
\label{plot_lya_lf}
\end{figure}

The best-fitting luminosity functions are plotted over the observed
abundances in the top panel of Figure \ref{plot_lya_lf} (simple model)
and in the bottom panel of Figure \ref{plot_lya_lf} (advanced model
with supernova feedback).  The error bars in these plots include both
Poisson and clustering variance.  Examining the confidence intervals
and luminosity functions, it seems that the data suggest some
evolution in the best-fitting model parameters between z=5.7 and
z=6.5.  It is unlikely that the evolution is associated with a change
in the neutral fraction of the IGM because the parameter that is
proportional to the transmission of Ly$\alpha$ photons, $\rm f_\star
T_{Ly\alpha}$, increases between z=5.7 and z=6.5, contrary to what
would be expected if the neutral fraction increased.  Taken at face
value, the evolution in the model parameters suggests that the star
formation efficiency and lifetime in LAEs increases between $z=5.7$
and $z=6.5$.  However, uncertainties in the observations make these
conclusions tentative.  The LAE luminosity function adopted in this
paper is based on a photometric sample of objects selected with a
narrowband filter.  While many of the most luminous
photometrically-selected objects have been confirmed
spectroscopically, at lower luminosities the completeness is still
low.  It is possible that there is significant contamination from
low-redshift line emitters, and hence that the densities at these low
luminosities are overestimated.  If the error bars at low luminosities
are enlarged to reflect this uncertainty, then \cite{Dijkstra2006}
claim  the model parameters are consistent with no evolution between
$z=5.7$ and $z=6.5$ except in  the underlying halos.  Additional
spectroscopic  observations of the lowest luminosity LAEs are clearly
necessary  to resolve whether the physical parameters of LAEs evolve
between  z=5.7 and z=6.5.

We now examine the consistency of our best-fit model parameters in the
context of additional observations of $z\gsim 5$ LAEs.  Recently,
observations have shown that many luminous LAEs at $z\simeq 5$ have
relatively low stellar masses.  In \cite{Pirzkal06}, LAEs at $z\simeq
5$ with L$\rm_{Lya} \simeq  10^{42} - 10^{43} erg s^{-1}$ have $\rm
M_{stellar} \simeq 10^6 - 10^8 ~  M_\odot$. Assuming the ratio of
baryons to  total mass follows the unversal value $\rm \Omega_b /
\Omega_m$  and a star formation efficiency of $\simeq 10$\%, the
observed  stellar masses suggest halo masses of $\rm 0.06 -
6\times10^9~M_\odot$.  According to our best-fit models, the halo
masses probed by the  Subaru observations are signficantly greater: at
$z=5.7$ halo masses  range from $\rm 10^{10}$ to $\rm 10^{11}
M_\odot$, while at $z=6.5$,  the halo masses range between $\rm
4\times10^{10}$ and $\rm 4\times$10$^{11}$  M$_\odot$. 

Given the low stellar masses observed for these objects, it seems that
one of our assumptions must be incorrect.  Under our simple model, the
stellar mass is given by $\rm M_{stellar} =
f_\star[\Omega_b/\Omega_m]M_{halo}$; in order to decrease the stellar
mass for a given halo mass, the star formation efficiency must be
lower than the 10\% value we have assumed above.  However, the star
formation efficiency must also satisfy the relation between Ly$\alpha$
luminosity and halo mass (equation 2) which is constrained via the
observed abundance of LAEs as a function of luminosity.  Hence, if the
star formation efficiency decreases, either the Ly$\alpha$
transmission factor or the ionizing photon rate must increase to
satisfy equation \ref{lya_model}.  The latter could occur if the IMF
of stars was more top-heavy than the standard Salpeter form.  While a
top-heavy IMF would reduce the stellar mass predicted by our model, it
would also decrease the observationally inferred LAE stellar masses.
These masses have been inferred via population synthesis models
assuming a Salpeter IMF.  Top-heavy IMFs have a lower rest-frame
optical stellar mass-to-light ratios then the Salpeter IMF does;
hence, for a given luminosity, the inferred mass in stars is lower
than for a Salpeter IMF.  Therefore, the stellar masses inferred from
observations remain at odds with those predicted from our models.
Alternatively, if the Ly$\alpha$ transmission fraction, T$_{\rm
  Ly\alpha}$, is enhanced to account for the lower star formation
efficiency, the models achieve much better agreement with the
observations.  At $\rm z=5.7$, our best-fitting models have $\rm
f_\star T_{Ly\alpha} \simeq 0.04$.  If the Ly$\alpha$ transmission
factor is near unity, then the star formation efficiency is roughly
4\%.  In this scenario, the stellar masses predicted by the models are
$\rm \simeq 10^7-10^8 ~ M_\odot$ at $z=5.7$, in much closer agreement
with the observations.

The low star formation efficiency and large Ly$\alpha$ transmission
factor of LAEs can be understood physically.  In addition to showing
that the brightest LAEs have low stellar masses, observations have
also shown that the most luminous LAEs are a young population with
ages of a few $\times10^6$ years \citep{Pirzkal06, Finkelstein06},
comparable to the lifetime of massive stars before they may explode as
supernovae.  Hence, these galaxies are observed at such a young stage
that they have not had sufficient time to convert more than roughly
4\% of their baryons to stars.  Moreover, they likely have not had
enough time to produce a significant amount of dust.  This conjecture
is corroborated by population synthesis modeling of the observed SEDs
of $z\simeq 5$ LAEs \citep{Gawiser06}.  Without dust to absorb the
resonantly scattered Ly$\alpha$ photons, the fraction of Ly$\alpha$
photons escaping the galaxy may increase substantially, explaining the
very large Ly$\alpha$ transmission factor needed to fit the observed
stellar masses. However, the Ly$\alpha$ transmission factor is also
dependent on intergalactic absorption.  The typical flux decrement
encountered by Ly$\alpha$ photons in the intergalactic medium
\citep{Fan06} may be substantially reduced in the vicinity of LAEs if
they reside in groups of galaxies which significantly ionize their
surroundings, allowing the Ly$\alpha$ photons to escape out of
resonance before encountering neutral hydrogen in the IGM
\citep{WLcl,Furlanetto06}.

\subsection{Comparison of LAEs and LBGs}

Recent observations at $z\simeq 5$ suggest that LAEs may differ  from
LBGs in their typical stellar mass and ages.  Observations  presented
in \cite{Eyles2006} have shown that $z\simeq 6$ LBGs  are a composite
population of galaxies, some with low stellar  masses ($\rm \simeq
10^8 ~ M_\odot$)  and some with high stellar masses ($\rm \simeq
10^{10} ~ M_\odot$).   These stellar masses emerge from our model
given the best-fit star  formation efficiency and halo masses probed
by the observations.  While the uncertainties are still significant,
there appears to  be a weak correlation between the stellar mass and
age of  $z\simeq 6$ LBGs \citep{Eyles2006}; the most massive galaxies
appear  to have a significant population of old stars (ages up to $\rm
\simeq 700 Myr$), while the less massive objects appear to be much
younger  (ages of $\rm \simeq 60~Myr$).  The model considered in this
paper fixes the  star formation duty cycle to be constant with halo
mass and hence cannot  confirm this observational inference.  The
best-fitting duty cycle for  LBGs suggests an age of 200 Myr, which is
roughly in the middle of the range  of lifetimes expected of the
LBGs. 

LAEs at $z\simeq 5$ also appear to be a composite population  spanning
a range of masses and ages, but it appears that their  typical ages
and stellar masses are systematically lower than LBGs.  A correlation
exists between the equivalent width (EW) of the  Ly$\alpha$ line and
the galaxy age and stellar mass \citep{Finkelstein06}.  The highest EW
lines exhibit the lowest ages (few million years) and  stellar masses
($\rm 10^6 - 10^7 ~ M_\odot$); these objects are very  faint continuum
emitters and hence are not selected in LBG surveys
\citep{Finkelstein06,Pirzkal06}.  LAEs with lower EWs have larger
inferred ages (40-200 Myr) with stellar masses up to $\rm  \simeq
10^{10}~M_\odot$ \citep{Finkelstein06, Lai07}; however, these ages and
masses do not reach the values as large as those seen in LBG surveys
at these redshifts.  Finally, the SEDs of the young LAEs at $z\simeq
5$  suggest that there is little exinction from dust in these galaxies
\citep{Pirzkal06}.  As explained in the \S4.2, in order for our  model
to fit the observations described above, the star formation efficiency
of the high EW LAEs must be low (due to their extreme youth)  and
transmission of Ly$\alpha$ photons through the host galaxy and  IGM
must be very high.

It is intriguing to consider the fate of the high EW LAEs.  In one
possible scenario, star formation continues, depleting the gas content
and increasing the total stellar mass of the galaxy.  The dust content
of the galaxy begins to increase. The dust absorbs the
resonantly-scattered Ly$\alpha$ photons, although the Ly$\alpha$ EW
might be enhanced depending on geometric details
\citep{Neufeld91,Peng}, If the star formation rate remains high, these
objects could continue to be observed as LBGs.  As more of the gas is
converted to stars, the star formation efficiency will increase.
Within our model we indeed inferred that the average star formation
efficiencies of the $z\simeq 6$ LBGs is on order 10\%.

Alternatively, the high EW LAEs could eject their gas via feedback
from supernovae explosions or quasar activity.  As soon as the gas
density is significantly diluted, the recombination rate decreases and
there is little emission of Ly$\alpha$ photons. The galaxies would
continue to be selected as LBGs as long as massive stars remain;
however, without gas star formation eventually ceases, leaving the
galaxies quiescent and with a low stellar mass.  Such objects would no
longer be detected as either LBGs or LAEs.

\section{Interpretations of Observations at $z\simeq 7-10$}

We now use our model fits, noting the uncertainties, to make
predictions for both LBGs and LAEs observed at $z>$7. We will assume
no evolution in the model parameters to determine what redshift trends
are expected solely from the natural growth of dark matter halos over
the era $5<z<10$. A modest amount of observational data is available
for the $z> 7$ universe and we will address this to see if it is
consistent with no evolution. Necessarily this discussion will be
tentative given the considerable uncertainty about the validity of the
observations. Most of the sources claimed to lie beyond $z\simeq 7$
have no convincing spectroscopic identification.

\subsection{Lyman-Break Galaxies}

Preliminary constraints are now available on the abundance of LBGs at
$z>7$ \citep{Bou05,Bou06b,Richard06}.  LBGs selected as z-band
dropouts are considered to have a mean redshift $z\simeq7.4$.  The
most recent compilation from fields with deep HST-NICMOS data
(e.g. GOODS, UDF, UDF-P) includes one candidate in the most
conservative selection and 4 candidates in a more aggressive selection
\citep{Bou06b}.  After comparing the observed abundance of candidate
$z\simeq 7.4$ LBGs to those at $z\simeq 6$, \cite{Bou06b} suggest that
there is a rapid assembly of the most luminous star forming systems
between $z\simeq 6$ and $z\simeq 7.4$.

By extrapolating our star-formation model to $z\simeq 7.4$ and holding
the parameters fixed at their best-fit $z\simeq 6$ values, we can
determine whether the claimed rapid assembly of luminous galaxies
requires evolution in the star formation efficiency or duty cycle.  If
all four candidate z-dropouts identified in \cite{Bou06b} are at
$z\simeq 7.4$, then the observed evolution in the abundance of
luminous galaxies can be explained simply by evolution of the host
dark matter haloes (Figure  \ref{plot_lbg_model}).  However, if only
one of the candidate z-dropouts  is at $z\simeq 7.4$, then our simple
model does permit some evolution  in either the star formation
efficiency or duty cycle in the 200 million years between $z\simeq
7.4$ and $z\simeq 6$.  Such evolution could be triggered, for example,
by the photo-ionization heating of the intergalactic medium at the end
of reionization and the corresponding change in its accretion rate
onto galaxies.

\begin{figure}
\figurenum{8} \epsscale{.95} \plotone{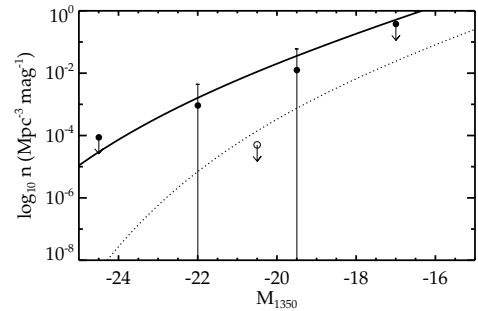}
\caption{Comparison of model $z=9$ LBG luminosity function with
  constraints from observations.  The observed abundance of LBGs
  presented in the cluster lensing survey of \cite{Richard06}, denoted
  by solid circles, is greater than the upper limit found in the
  Hubble Ultra Deep Field (open circle), as presented in \cite{Bou06}.
  The error bars on the  \cite{Richard06} data are large but may
  overestimate the true uncertainty (see \S 5.1). The luminosity
  function obtained by assuming the duty cycle and star formation
  efficiency remain constant between $z=6$ and $z=9$ (dotted line) is
  underpredicts the large abundances observed in \cite{Richard06}.
  Reconciling these data points with our models without resorting to
  more top-heavy IMFs require a star formation efficiency near 100\%
  (solid line).  }
\label{plot_lbg_z9}
\end{figure}

At $z\simeq 10$, several candidate LBGs (selected as J-band dropouts)
have been identified in the UDF \citep{Bou05} and in cluster lensing
fields \citep{Richard06}.  With the addition of deeper optical data,
two of the three candidate $z\simeq 10$ LBGs from \cite{Bou05} are now
known to be at lower-redshift (R.J. Bouwens 2006, private
communication). In this case, the abundance implied if the remaining
one candidate is at $z\simeq 10$ is consistent with hierarchical
growth (Figure \ref{plot_lbg_z9}).  

In contrast, the abundance of less luminous $z\simeq 9$ candidates
located in \cite{Richard06} is significantly larger than expected
(Figure \ref{plot_lbg_z9}).  However, if clustering fluctuations are
included in the uncertainties, then the abundances derived are
formally consistent with the lower density implied by \cite{Bou05}.
We note that the 1-sigma error bars on the \cite{Richard06} data
include a large contribution (typically a factor of 3) from the
uncertainty in the completeness correction.  This uncertainty is
difficult to determine accurately; if overestimated by as little as
$\simeq 20$\%, the one-sigma uncertainties on the \cite{Richard06}
datapoints would no longer be consistent with the \cite{Bou05}
observations.  More clusters must be studied to verify the large
density of lensed $z\simeq 9$ candidate LBGs.  If the large abundances
are representative, the \cite{Richard06} observations require either
significant evolution in the parameters of the simple model we have
adopted or a top-heavy stellar IMF.  Holding the duty cycle fixed at
its $z\simeq 6$ value, an implausible star formation efficiency of
100\% is required to explain the observed abundance (Figure
\ref{plot_lbg_z9}).  

The emerging physical picture describing the evolution of LBGs at
$z\simeq 7-10$ is still somewhat tentative. Nonetheless, results from
the UDF suggest  that evolution in the abundance of {\it luminous}
LBGs in the 500 Myr between  $z\simeq 6$--$10$ can be largely
explained by the hierarchical assembly, i.e. without any evolution in
the star formation efficiency or  duty cycle.  In contrast, the high
abundance of {\it less luminous}  $z\simeq 10$  candidates suggested
by \cite{Richard06} may require significant evolution in either the
stellar IMF or the star formation efficiency.  Similar conclusions are
reached using semi-analytic models in \cite{Samui06}.  Additional
observations are required to confirm the large density observed in
\cite{Richard06} is robust  and to reconcile these potentially
differing pictures.  In $\S6$, we will use our model to predict the
ability of future surveys to detect starburst galaxies at  $z\simeq
7-10$.

\subsection{Ly$\alpha$ Emitters}

The first results are also now available from Ly$\alpha$ surveys at
$z\simeq 9$.  As with the LBGs, the high redshift results are
seemingly  contradictory.  Willis et al. (2005) and Cuby et al. (2006)
find no Ly$\alpha$ emitters in narrowband surveys centered at $z=8.8$
with the Very Large Telescope (VLT).   However, both surveys are only
sensitive to the brightest LAEs  ($>3\times10^{42}$ erg~s$^{-1}$ in
\citealt{Willis2005} and $>$10$^{43}$ erg~s$^{-1}$ in
\citealt{Cuby2006}) over modest comoving volumes (870 Mpc$^3$ and 5000
Mpc$^3$, respectively).  

\cite{Stark07b} conducted a cluster lensing survey for LAEs at
$z=8.5-10.4$.  The magnification provided by the clusters allows
significantly less luminous LAEs to be detected ($\gsim$10$^{41}$ erg
s$^{-1}$), albeit over a much smaller comoving volume ($\simeq$ 30
Mpc$^3$).  Six candidate LAEs were identified with unlensed
luminosities spanning 2--50$\times$10$^{41}$ erg s$^{-1}$; at least
two of  the six candidates are considered likely to be at $z\simeq 9$
following additional spectroscopy which casts doubt on alternative
low-redshift explanations for the J-band emission features.

\begin{figure}
\figurenum{9} \epsscale{0.95} \plotone{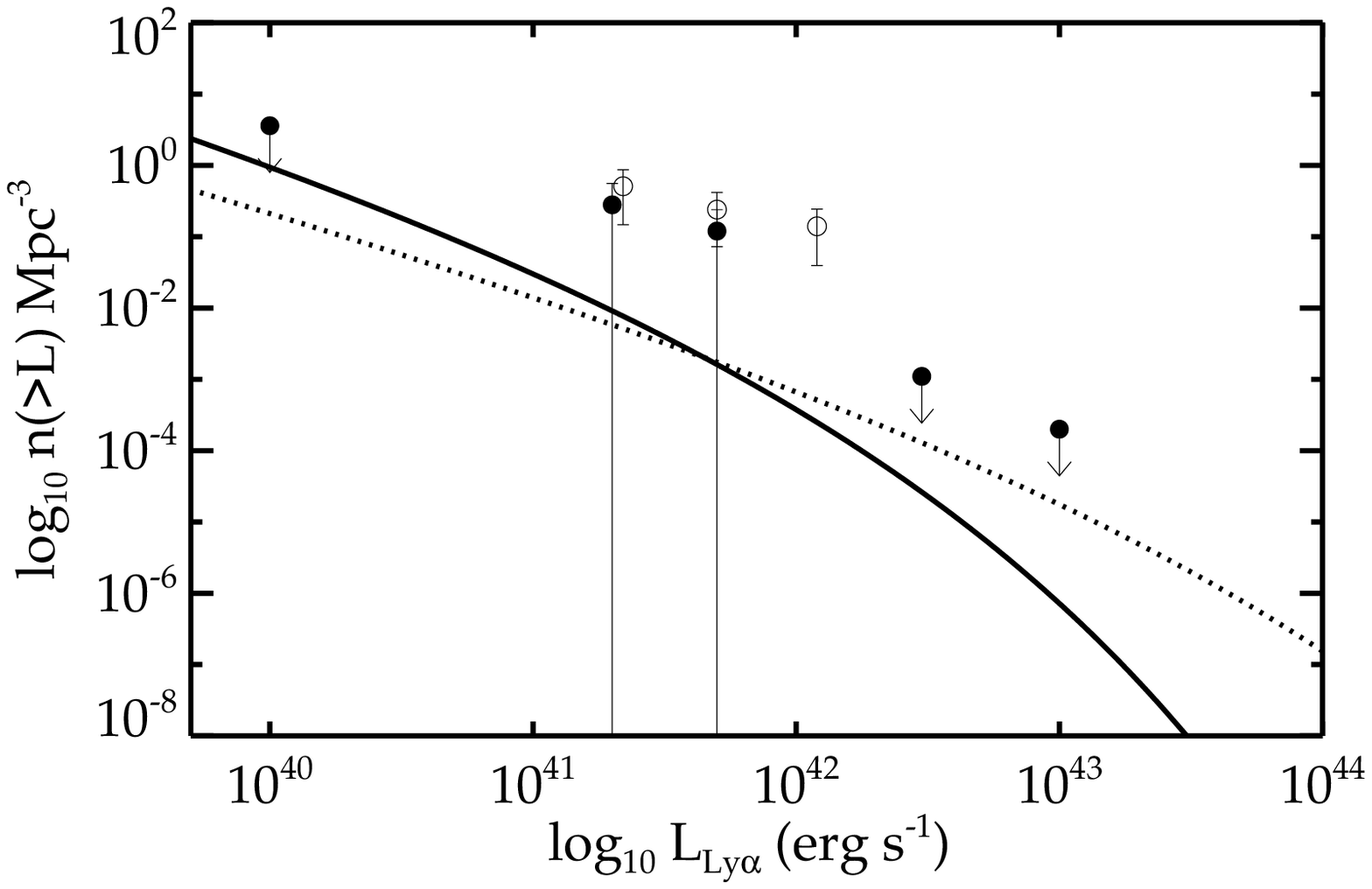} \plotone{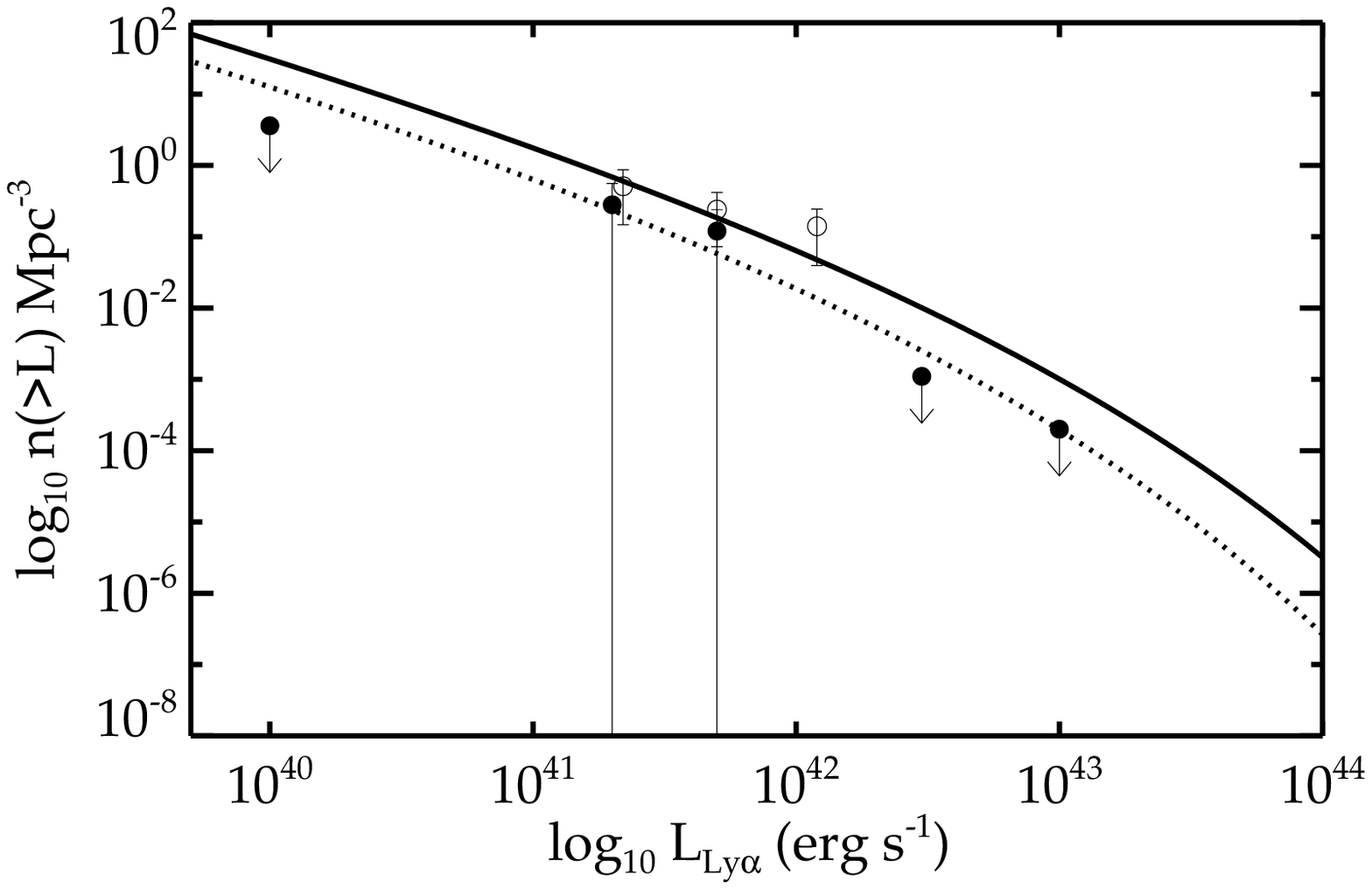}
\caption{Comparison of model $z\simeq 9$ LAE luminosity function with
  constraints from observations.  {\it Top:} If all the $z\simeq 9$
  LAE candidates in \cite{Stark07b} are at high-redshift, then the
  best-fit model parameters at $z=5.7$ (dotted line) and $z=6.5$
  (solid line) are inconsistent with their values at $z=9$.  {\it
    Bottom:} Model luminosity functions assuming an extreme top-heavy
  IMF characteristic of Pop-III stars with model parameters fixed at
  their $z=6.5$ (dotted line) and a star formation efficiency of 100\%
  (with duty cycle fixed at its best-fit z=6.5 value, solid line)
  overlaid upon observational constraints.}
\label{plot_lya_z9}
\end{figure}

In Figure \ref{plot_lya_z9}, we use our model to compute luminosity
functions of LAEs at $z\simeq 9$ assuming the duty cycle, star
formation  efficiency, and Ly$\alpha$ escape fraction remain fixed at
either their  best-fit $z=5.7$ or $z=6.5$ values.  As the narrowband
observations refer to a single redshift, we use the mass function at
$z=8.8$ to compute the luminosity function. For the cluster lensing
survey presented by  \cite{Stark07b}, as the halo mass function
evolves significantly over the redshift range sampled, we compute the
average halo mass function  between $z=8.5$ and $z=10.4$, weighting
mass by the relevant sensitivity function.  We generated luminosity
functions using both mass  functions. While the resulting luminosity
functions are marginally different,  the net results described do not
change; hence, for the sake of clarity,  in Figure \ref{plot_lya_z9}
we only overlay the luminosity function from the $z=8.8$ mass function
on the data  points from the three surveys described above.

The results suggest that, for {\it luminous} LAEs, current surveys do
not yet have the combined sensitivity and depth to detect any sources
at $z\simeq 9$.  Although the upper limits presented in
\cite{Willis2005} and \cite{Cuby2006} are consistent with our
expectations, those surveys only  rule out the possibility that the
density of luminous LAEs decreases in the  time interval between
$z\simeq 9$ and $z\simeq 6$.  On the other hand,  for {\it less
  luminous LAEs}, if all six of the candidates in  \cite{Stark07b} are
at high-redshift and the inferred abundances are representative,
significant evolution is implied in the model parameters.   As with
the lensed LBGs at $z\simeq 9$, either a high ($\simeq$ 100\%) star
formation efficiency (for a fixed duty cycle) or a top-heavy IMF of
stars (Figure \ref{plot_lya_z9}) would be required.  An important
caveat is the uncertainty caused by cosmic variance, which we do not
include in the error bars in Figure \ref{plot_lya_z9}.  Even for a
significantly more ambitious spectroscopic lensing survey, the
fluctuations expected from large-scale structure are $\gsim 100$\%
(Figure \ref{plot_sigma_slit}); hence, it is possible that the
candidate LAEs discovered in \cite{Stark07b} may trace an extreme
overdensity in the underlying mass distribution, in which case the
derived densities may be larger than the cosmic average at that epoch.
Clearly, more clusters must be observed.  If only the two prime LAE
candidates are at high redshift\footnote{The fact that the Poisson
  errors are 100\% for the two datapoints in Figure \ref{plot_lya_z9}
  corresponding to the case that two candidates from \cite{Stark07b}
  are at $z\simeq 10$ may seem counterintuitive given that there are
  two sources; however this arises from strongly-varying limiting
  (detectable) source luminosity over the field of view due to the
  cluster magnification.  For details, see \cite{Stark07b}.}, the
derived abundances are formally consistent with the predicted $z\simeq
9$ luminosity function due to the large Poisson fluctuations.

\section{Implications for Future Surveys}

Over the next few years, several new instruments will be placed on
current ground-based telescopes and Hubble Space Telescope motivated
in part by extending the search for $z>$7 sources. These surveys offer
the exciting possibility  of characterizing the assembly of the first
galactic sources through  the era of reionization.  With our model,
calibrated by the data  available at $z\simeq 6$, we can consider the
optimal volume and  depth necessary for detecting star-forming sources
of various luminosities to $z\simeq 10-20$. Thus it is hoped our model
can assist in guiding the  design of future instruments and surveys. 

In the following subsections, we will evaluate several benchmark
imaging and spectroscopic surveys.  We consider a large, blank-field,
narrowband survey for LAEs at  $z\simeq 8$ and $z\simeq 10$ in \S 6.1,
and discuss two different surveys for  dropouts at $z\simeq 7.5$ and
$z\simeq 10$ in \S 6.2.  Finally, in \S 6.3, we  examine the
efficiency of a variety of lensing surveys for LAEs and LBGs.  Since
some of these proposed surveys reach to significantly lower
luminosities than the current surveys discussed in \S4 and \S5, we
examine the effects of feedback on the predicted counts, as well as
the gains of adaptive optics given our predicted sizes for faint
LAEs. 

\subsection{The Dark Ages z Lyman-alpha Exlorer: LAEs at $z\simeq 7-10$}

The Dark Ages z Lyman-alpha Explorer (DAzLE) is a narrowband imager on
the VLT which aims to detect Ly$\alpha$ emitters at $6.5<z<12$
\citep{Horton04}.  DAzLE has recently observed two pointings of
GOODS-South  in two filters corresponding to Ly$\alpha$ redshfits of
$\simeq 7.7$  and $\simeq 8.0$ (R. McMahon 2006, private
communication).  The observing  sequence for DAzLE involves
alternating between two narrowband filters with slightly  different
central wavelengths.  A composite image  is made of all of the
subexposures in each filter, resulting  in two ``subsurveys'' slightly
offset in redshift space. Subtracting  the two composite images
removes continuum soures, thereby allowing  candidate LAEs to be
identified.   

In ten hours of integration, DAzLE is expected to reach a 3$\sigma$
sensitivity of $2\times10^{-18}$ erg cm$^{-2}$ s$^{-1}$ in the
differenced image \citep{Horton04}.  At $z=7.7$, this corresponds to a
limiting LAE luminosity of $1.5\times10^{42}$ erg s$^{-1}$.  As a
benchmark survey with this instrument, we consider a four position
mosaic (i.e. 4$\times$ 6.\arcmin83$\times$6.\arcmin83). The total
comoving volume sampled in four pointings of the two filters at $z =
7.7$ is $\rm\simeq 6900~Mpc^3$.  A simple extrapolation of our model
suggests that a comoving volume of 1100 Mpc$^3$ is necessary to detect
one LAE at $z=7.7$.  Thus, in this proposed survey, 6-7 LAEs would be
detected with DAzLE assuming no rapid evolution in the star formation
efficiency, duty cycle or IGM transmission between $z\simeq 6.5$ and
$z\simeq 7.7$.

However, although a promising survey in terms of likely detections,
the uncertainties are considerable. Clustering fluctuations are
40-50\% (for the best-fit z=5.7 and 6.5 model parameters,
respectively) at the limiting luminosity of $1.5\times10^{42}~{\rm
  erg~s^{-1}}$.  The Poisson fluctuations expected for a 7200 Mpc$^3$
survey are similar ($\simeq 40$\%).  It is prudent to consider what
effects such large fluctuations would have on attempts at using the
DAzLE results to constrain the progress of reionization via the
evolution of the LAE luminosity function
\citep{Malhotra2004,Malhotra2006,Dijkstra2006}.  While our model
predicts that  6-7 sources should be detected assuming only evolution
in the underlying halos, the large ($\simeq 60$\%)  expected
field-to-field variations imply that the source counts could vary
significantly from the predicted value.  Ignoring additional
complications on the tranmission of Ly$\alpha$ photons through the IGM
from galaxy groups \citep{WLcl,Furlanetto04,Furlanetto06} and peculiar
velocities (Dijkstra \& Loeb 2006, in preparation), if between two and
ten LAEs are detected with DAzLE in GOODS-S, little information can be
reliably deduced on the evolution of IGM.  Holding the duty cycle and
star formation efficiency fixed, it would require a $\gsim $60\%
decrease in the Ly$\alpha$ transmission factor, T$_\alpha$, for only
one $z=7.7$ LAE to be detected towards GOODS-S.  Hence, these results
suggest that, with currently feasible survey geometries and
instruments, this method of constraining reionization will only be
effective if the IGM evolves rapidly ($\gsim $60\%) over a short
redshift interval.

\begin{figure}
\figurenum{10} \epsscale{1.} \plotone{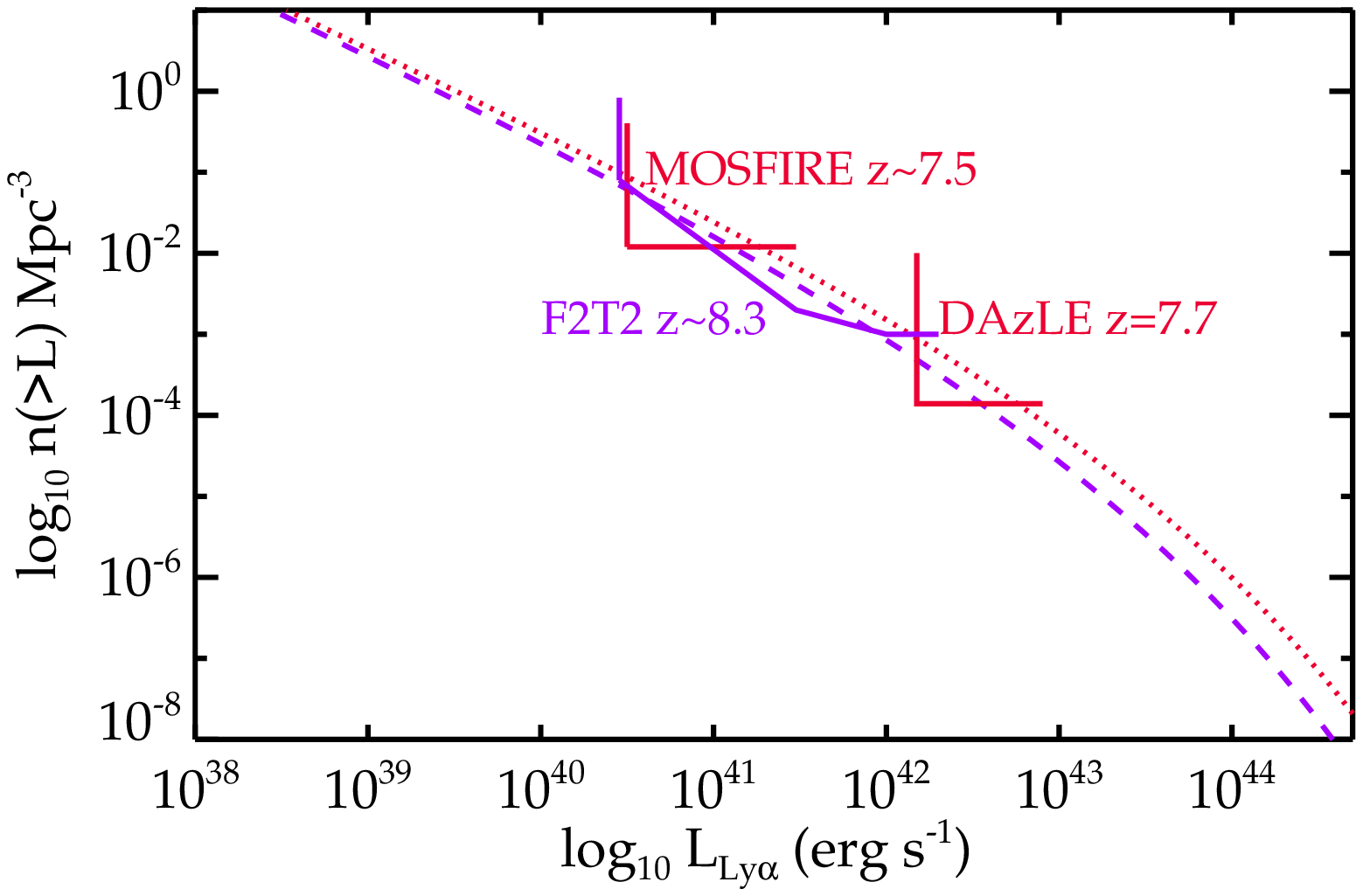} \plotone{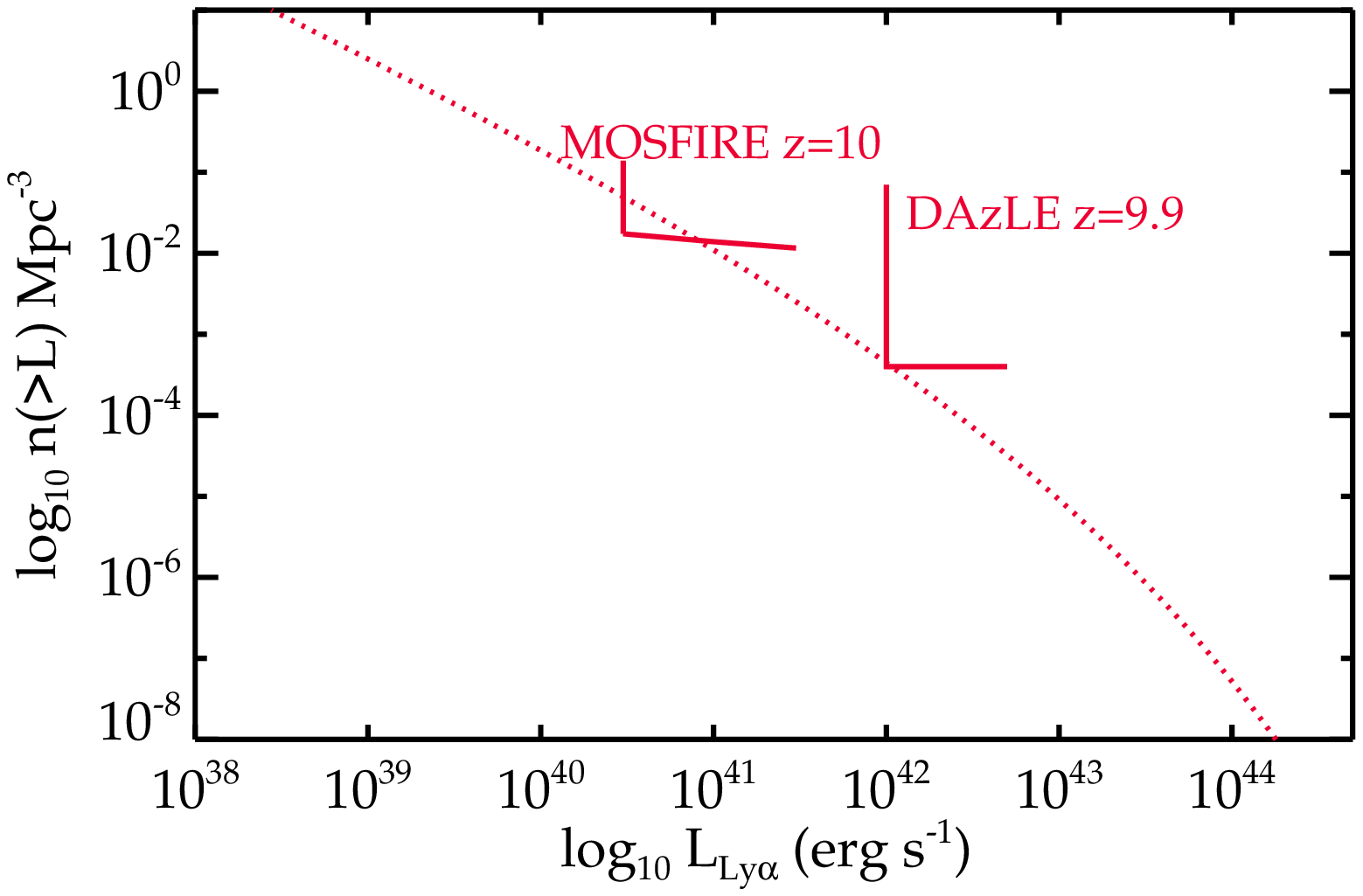}
\caption{{\it Top:} Extrapolation of Ly$\alpha$ luminosity function to
  $z\simeq $ 7.7 (dotted red curve) and 8.3 (dashed purple curve)
  fixing star formation model parameters at their best-fit $z=6.5$
  values (\S 4).  The predicted luminosity function suggests that
  DAzLE should detect 6-7 $z\simeq 7.7$ sources in a four pointing
  mosaic with total integration time of $\simeq 40$ hours.  A lensing
  survey utitlizing F2T2 could  detect up to 8 $z\simeq 8.3$ sources
  in $\simeq 200$ hours of observations.  {\it Bottom:}  Extrapolation
  of Ly$\alpha$ luminosity function to $z\simeq $ 10 fixing star
  formation model parameters at their best-fit $z=6.5$ values.  Almost
  200 hours are required to detect a single $z\simeq 10$ source with
  DAzLE under these assumptions.  The estimated performance of lensing
  surveys with a multi-object spectrograph such as MOSFIRE or a
  tunable narrowband filter (F2T2) suggest that lensing is an
  efficient means of detecting LAEs at $z\simeq 7-10$, but supernova
  feedback could drastically reduce the number of LAEs detected. }
\label{plot_lya_z7-10}
\end{figure}

DAzLE is able to search for LAEs out to $z\simeq 10$.  There is a
large gap in the atmospheric OH forest between 1.3325 $\mu$m and
1.3401 $\mu$m, corresponding to a LAE redshift window of $\Delta
z$=0.06 centered at $z=9.99$.  A similar window is located at
$z=9.91$.   A single DAzLE pointing at this redshift (assuming a
filter  width of $\simeq 10$\AA) samples a comoving volume of $\simeq
1250~{\rm Mpc^3}$ summed over both filters.  Extrapolating the model
luminosity function to this redshift with the parameters fixed at
their best-fit $z=5.7$ values, it would take just over a volume of
2500 Mpc$^3$ (two pointings) to detect one LAE brighter than 10$^{42}$
erg s$^{-1}$.  At $z=9.99$, this luminosity limit corresponds to a
limiting flux of $\rm 8\times10^{-19}~erg~cm^{-2}~s^{-1}$.  Reaching
this sensitivity would require 70--80 hours for each pointing
(adjusting for the expected atmospheric transparancy in the wavelength
interval of the observations); hence almost 200 hours would be
required to detect a single $z\simeq 10$ source.  Alternatively, if
the LAE candidates identified by \cite{Stark07b} are at high-redshift,
then $\gsim 100$ LAEs brighter than 10$^{42}$ erg s$^{-1}$ would be
expected in the single 70-80 hour DAzLE pointing.

In summary, DAzLE may well detect many $z$=7.7 sources but
spectroscopic confirmation will be a challenge.  However, even in the
most ambitious surveys we can currently contemplate, the contribution
of these sources to reionization will be seriously limited by the
expected clustering fluctuations.  Surveys at higher redshift will be
much more demanding. Few sources are expected within reasonable
exposure times at $z\simeq 9$--$10$ unless there is significant
evolution in the LAE model parameters between 6$<z<$10, as may be
possible only if all of the fainter lensed LAEs detected by
\cite{Stark07b} are at high redshift.

\subsection{Imaging Surveys for LBGs at $z\simeq 7-10$}

Selection of $z\gsim 7$ LBGs will be greatly aided by the new
generation of large-format near-infrared detectors.  The Subaru
Multi-Object Infrared Camera and Spectrograph (MOIRCS,
\citealt{Ichikawa06}) offers imaging and  spectroscopic capabilities
over a large $4\arcmin\times7\arcmin$ field of view.  In 2008, the
Wide Field Camera 3 (WFC3) is scheduled to be installed on HST; this
near-IR camera will offer imaging in a 127\arcsec $\times$137\arcsec ~
field of view.  We thus consider likely surveys with MOIRCS and WFC3.
First we consider a MOIRCS wide-field mosaic whose sensitivity is
arranged to match that of the near-infrared observations of GOODS-S,
but with an area twice as large.  We then consider a single,
ultra-deep pointing with WFC3.  

To observe an area twice as large of GOODS-S would require eleven
MOIRCS pointings.  Selecting reliable z-dropouts ($z\simeq 7.5$)  and
J-dropouts ($z\simeq 10$) requires deep z, J, H, and K-band data.  We
assume the near-infrared data is roughly similar in depth to GOODS,
with 5-$\sigma$ point sources sensitivies of $z'$=26.6, J=25.8, H=24,
and K=24.  These limits are optimized to the selection criteria of
$z'$-drops ($z'-J>$0.8, \citealt{Bou06b}) and J-drops (J-H$>$1.8,
\citealt{Bou05}).  The K-band limit is chosen with the goal of
detecting J-drops in a second filter to help remove false-positives.
The limiting J- and H-band limits correpond to limiting star formation
rates of 18 and 146 $\rm M_\odot yr^{-1}$ at $z\simeq$ 7.5 and 10,
respectively.  Reaching J=25.8 with MOIRCS requires $\simeq 12$ hours
of integration assuming 0\farcs5 seeing (J. Richard 2006, private
communication).  Significantly less time is needed to reach the
desired sensitivies in H (1.8 hours) and K (40 minutes).  The
$z'$-band observations would be most efficiently performed with the
Suprime-Cam on Subaru.  The entire area could be covered in one
pointing with Suprime-Cam, requiring 4.7 hours of integration.  In
total, 163 hours would be required for the observing sequence. We
consider this a practical, albeit ambitious, program.

\begin{figure}
\figurenum{11} \epsscale{0.95} \plotone{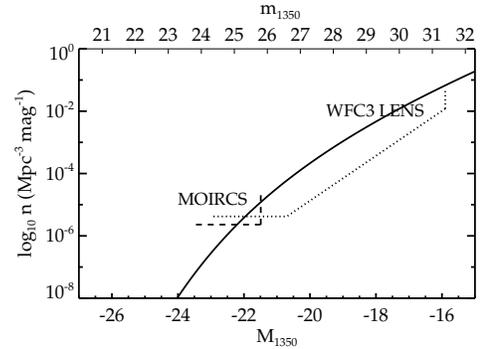}
\caption{Predicted LBG luminosity function at $z\simeq 7.5$ assuming
  star formation model parameters are fixed at their best-fit $z\simeq
  6$ values (\S 4).  The absolute magnitude at 1350\AA ~is plotted
  horizontally along the bottom of the plot, and the corresponding
  J-band  apparent magnitude is plotted along the top of the plot,
  assuming  $z=7.5$.  The efficiency of several mock surveys for
  $z\simeq 7.5$ sources is overplotted.   }
\label{plot_lbg_z7.5}
\end{figure}

Each MOIRCS field of view samples a comoving area of 10.6 Mpc $\times$
18.6 Mpc at $z\simeq 7.5$ and 11.4 Mpc $\times$ 20.0 Mpc at $z\simeq
10$.   We assume the redshift probability distribution of z-drops and
J-drops is a Gaussian with a standard deviation of $\sigma_z$=0.5 and
a mean of $\mu_z$=7.5 and 10 for z-drops and J-drops, respectively.
We normalize the probability distribution so that the maximum
completeness  is 50\% at the mean redshift of the survey. While the
completeness is  certainly higher for objects that are much brighter
than the sensitivity  limit of the survey, monte carlo simulations
suggest that 50\% is a reasonable average for entire population which
is  dominated by the faintest objects  \citep{Stark07a}.  We compute
the effective radial distance sampled by the surveys by integrating
the normalized redshift probability distribution over a distance
spanning $\Delta z$=2 in redshift and centered at $\mu_z$.  Over
eleven pointings, this corresponds to a total comoving volume of
4.4$\times 10^5 ~ {\rm Mpc^3}$ at $z\simeq 7.5$ and 3.5$\times10^5
~{\rm Mpc^3}$ at $z\simeq 10$.

\begin{figure}
\figurenum{12}  \epsscale{1.1} \plotone{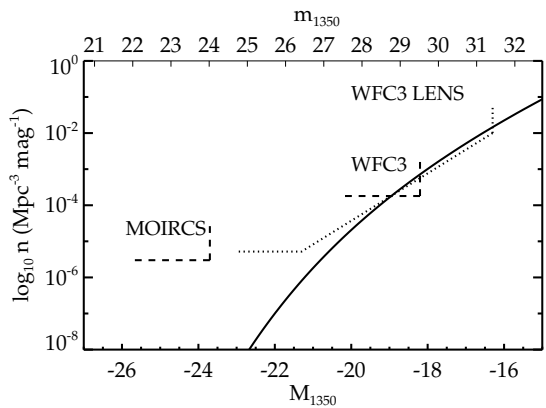}
\caption{ Predicted LBG luminosity function at $z\simeq 10$ assuming
  star formation model parameters are fixed at their best-fit $z\simeq
  6$ values (\S 4).  The absolute magnitude at 1350 \AA\ is plotted
  horizontally along the bottom of the plot, and the corresponding
  H-band  apparent magnitude is plotted along the top of the plot,
  assuming  $z=10$.  The efficiency of several mock surveys for
  $z\simeq 10$ sources is overplotted.   }
\label{plot_lbg_z10}
\end{figure}

In Figures \ref{plot_lbg_z7.5} and \ref{plot_lbg_z10}, we extrapolate
the best-fit luminosity function from $z\simeq 6$ to $z\simeq 7.5$ and
$z\simeq 10$ allowing only for evolution in the dark matter mass
function. The comoving number density of detected LBGs at the
sensitivity limit of  the MOIRCS survey is $1.2\times10^{-5}$
Mpc$^{-3}$ mag$^{-1}$ at $z\simeq 7.5$ and $1.3\times10^{-10}$
Mpc$^{-3}$ mag$^{-1}$ at $z\simeq 10$.  Integrating over the entire
magnitude range, 1-2 sources would be detected brighter than J=25.8 at
$z\simeq 7.5$ and 4$\times10^{-6}$ sources would be detected brighter
than H=24 at $z\simeq 10$.  Clearly, neither the $z\simeq 7.5$ or
$z\simeq 10$ MOIRCS mock-survey is very efficient at detecting
high-redshift sources.  Even with the relatively large areal coverage
provided by MOIRCS, the integrated near-infrared sky is simply too
bright in broadband filters to reach the sensitivity limits necessary
to detect an abundant population of $z\gsim 7-10$ LBGs; with current
technology, conventional ground-based surveys are much better off
searching for LAEs using narrower filters tuned in between the bright
sky lines.

As with the MOIRCS observations, the mock WFC3 observations  require
deep z, J, and H-band coverage to select z- and  J-band dropouts.
WFC3 offers a significant improvement in both  throughput and areal
coverage with respect to NICMOS, so the camera  is ideal for selecting
J-drops.  By contrast,  WFC3 is {\it not} more efficient than ACS in
the z-band.  Therefore, WFC3  is not particularly well-suited for
conducting a new z-drop survey that  is significantly deeper than UDF.
We thus consider its efficiency at detecting J-dropouts.

We adopt limiting magnitudes that are $\sim$1 and $\sim$2 magnitudes
deeper than the UDF in $\rm H_{160}$ and $\rm J_{110}$, respectively,
corresponding  to 5$\sigma$ sensitivities of $\rm J_{110}$=31.3 and
$\rm H_{160}$=29.5 for point sources (assuming an 0\farcs4 diameter
aperture). With a J-band 5$\sigma$ sensitivity of 31.3, J-drops can be
selected as faint as $\rm H_{160}$=29.5, using the selection criteria
of \cite{Bou05}.  The estimated star formation rate for $z\simeq 10$
LBGs with $\rm H_{160}$=29.5 is 0.9 M$_\odot \rm yr^{-1}$. Based on
the anticipated WFC3 performance, reaching such sensitivities in a
single pointing would take 301 hours in $\rm J_{110}$ and 48 hours in
$\rm H_{160}$ resulting in 349 hours of total
integration \footnote{Estimated sensitivies for WFC3 are listed at
  {\tt http://www.stsci.edu/hst/wfc3/documents/handbook/
    cycle16/wfc3\_cyc166.html}}. Since K-band observations  are not
practical with WFC3, we rely only on an H-band detection for J-drop
selection.  Each WFC3 field of view covers a comoving area of  6.0 Mpc
$\times$ 6.5 Mpc at $z\simeq 10$.  Assuming a redshift probability
distribution identical to that described above for z-drops and J-drops
with MOIRCS, we find that the single WFC3 pointing  will sample a
comoving volume of 5410 Mpc$^3$ at $z\simeq 10$. 

The predicted comoving number density of detected J-drop LBGs at the
sensitivity  limit of the single deep WFC3 pointing is $7.2\times
10^{-4}~{\rm Mpc^{-3}~mag^{-1}}$ at $z\simeq 10$, assuming
no-evolution in the star formation efficiency and duty cycle.
Integrating over the entire magnitude range, 1-2 sources would be
detected at $z\simeq 10$.

In summary, existing ground-based imagers are not well-equipped for
detecting LBGs at $z\simeq 7-10$, largely because the integrated
near-IR sky is too bright to reach the sensitivity limits necessary to
detect sources at $z\gsim 7$ in a reasonable amount of time.  This
problem could be lessened if the field of view was much bigger,
allowing the the most luminous and rare LBGs to be detected.  Even
from space (using WFC3 on HST), detecting LBGs at $z\simeq 7-10$ will
not be trivial, requiring hundreds of hours to detect a single LBG at
$z\simeq 10$.

\subsection{Lensing Surveys for Star-forming Galaxies at $z\simeq 7-10$}

Currently, strong lensing surveys for high-redshift galaxies  could be
contemplated for about 20 galaxy clusters for which there are
well-defined mass models; these mass models are essential in
accurately defining the spatial distribution of magnification.
However, ongoing HST surveys will significantly increase the  number
of suitable galaxy clusters (e.g. \citealt{Ebeling03}).  Here,  we
examine the efficiency of surveys for lensed LBGs and  LAEs assuming a
larger sample of clusters and more efficient  instruments that will
soon become available.  We consider both an  extension of the longslit
spectroscopic survey discussed in  \citep{Stark07b} and as well as an
imaging campaign  to identify lensed z-drops and J-drops.

In the next several years, a number of near-IR multi-object
spectrometers will be installed on 8-10 meter class telescopes,
offering  significant gains in sensitivity and field-of-view over
current  near-IR instruments.  One such instrument is the Multi-Object
Spectrograph  For Infra-Red Exploration (MOSFIRE, PI: I. McLean \&
C. Steidel) on Keck I.  MOSFIRE will utilize a configurable slit unit
allowing up to 45 slits, each  7\farcs3 in length, within the
6.\arcmin14$\times$6.\arcmin14 field of view.  While tilted and curved
slits are not possible with MOIRCS, the  configurable slit unit will
still allow for more areal coverage  of high magnification regions
than with NIRSPEC.  In addition,  MOSFIRE will be more sensitive than
NIRSPEC, reaching 5$\sigma$  line flux sensitivities of
1$\times$10$^{-18}$ erg cm$^{-2}$ s$^{-1}$  in 4 hours, assuming an
unresolved line with R=3270 \footnote{Sensitivities  and additional
  specifications of the spectrograph are provided in the  Preliminary
  Design Report located at {\tt http://www.astro.ucla.edu/~irlab/
    mosfire/MOSFIRE\%20PDR\%20Report\%20v4.pdf}}.  Given the
specifications of MOSFIRE, it is optimal to  concentrate observations
to clusters with vary large and well-determined  critical lines
(e.g. Abell 1689, Abell 1703, Abell 2218).  While the exact
magnification distribution provided to background sources may very
slightly from  cluster to cluster, it is reasonable to expect 80\% of
the MOSFIRE survey area to be at a magnification of 20, 15\% of the
area to have a magnification  of 10, and 5\% of the area to be
magnified by only a factor of 5  \citep{Stark07b}.  

We consider a MOSFIRE {\it spectroscopic} lensing survey for LAEs both
at $z=7.0-8.3$ (Y-band) and at $z=8.5-10.3$ (J-band).   Assuming
$\simeq 15$\% of the line-of-sight  distance is lost due to bright OH
lines, this corresponds to comoving radial distances of 353 and 334
Mpc, respectively.  Assuming 10 clusters are observed for 8 hours each
in the Y-band, the resulting  survey volume is $\rm\simeq 73~ Mpc^3$
for LAEs brighter than $\rm 10^{40.5} erg s^{-1}$.  For the mock
J-band survey, we assume longer integration times (20 hours) over
eight clusters, giving an identical integration time to the mock
DAzLE $z=9.9$  survey considered previously.  This results in the
coverage of $\rm\simeq 58~ Mpc^3$ for $z\simeq 10$ LAEs brighter than
$\rm 10^{40.5} erg s^{-1}$.  Extrapolating our LAE model to  $z=7.5$
and to $z=10$ holding the model parameters fixed at their $z=5.7$
values, we predict that the MOSFIRE survey should detect 9 sources
brighter than $\rm 10^{40.5} erg s^{-1}$  at $z=7.5$ and 4 sources
brighter than this limit at $z\simeq 10$.  If supernova feedback
decreases the star formation efficiency in a manner described in \S 2,
then the predicted number of LAEs would drop  drastically.  The
$z\simeq 7.5$ survey would potentially detect 1-2 LAE brighter than
$\rm 10^{40.5} erg s^{-1}$ while the $z\simeq 10$ survey  would not
detect any sources.  Hence, the success of the mock lensing survey
depends strongly on whether supernova feedback decreases the
efficiency of  star formation in low-mass dark matter halos.  Given
the possible efficiency  with which the lensing survey could detect
$z\simeq 7-10$ galaxies, it is of the utmost importance to
observationally constrain the effects of supernova feedback on the LAE
luminosity function at lower redshifts ($z\simeq 3-6$).

Gravitational lensing can also be very valuable for {\it imaging}
surveys for LBGs that are intrinsically fainter than those  detected
in conventional deep surveys (e.g. GOODS, UDF).    One such survey is
currently being conducted toward six galaxy clusters with NICMOS on
HST (\citealt{Stark06},  Richard et al. 2007, in preparation).  With
WFC3, such a survey could potentially be conducted much more
efficiently, allowing many more clusters to be observed.  We examine
the feasibility of a hypothetical WFC3 lensing survey of galaxy
clusters for z- and J-dropout galaxies.  The primary benefit of a WFC3
survey for $z\gsim 7$ LBGs is the added throughput and field of view
in the J$_{110}$ and  H$_{160}$-bands compared to what is available
with NICMOS. For each cluster, we assume 5-$\sigma$ point-source
sensitivities  (in an aperture with 0\farcs4 diameter) of
z$_{850}$=27.4, J$_{110}$=28.2,  and H$_{160}$=27.4, requiring $\simeq
8$, 1, and 1 hour(s) per cluster,  respectively.   These limits allow
z-drops to be selected down to J$_{110}$=26.6 and J-drops to be
selected down to  H$_{160}$=26.4 (without considering the effects of
lensing).  If we allot 350 hours to this observing program (an
identical  time allocation to the traditional survey discussed in the
previous subsection), 35 clusters can be observed.  The total area
surveyed would be more than a factor of fifteen greater than previous
lensing surveys for $z\gsim 7.5$ LBGs (Richard et al. 2007, in
preparation).

Both the survey volume and limiting source luminosity are modified by
the magnification provided by the foreground galaxy cluster (see
\citealt{Santos04} or \citealt{Stark07b} for a description).  A
typical cluster provides a magnification boost of $\times$2,5,10, and
30 over 92\%, 46\%, 29\%, and 12\% of the entire WFC3 field of view.
Adopting this magnification distribution for each of the 20 clusters,
at $z\simeq 7.5$, the survey is sensitive to a volume of
2.4$\times10^5$ Mpc$^3$, 7.6$\times10^4$ Mpc$^3$, 1.5$\times$10$^4$
Mpc$^3$, 3.1$\times$10$^3$ Mpc$^3$, and 580 Mpc$^3$ for sources
brighter than J$_{110}$ = 26.6, 27.6, 28.6, 29.6,  and 30.6
respectively\footnote{The quoted magnitudes correspond to the apparent
  magnitude that a source would be observed with if it was not
  magnified.}.  At $z\simeq 10$, the survey probes a comoving volume
of 1.9$\times10^5$ Mpc$^3$, 6.0$\times10^4$ Mpc$^3$, 1.1$\times10^4$
Mpc$^3$, 2.4$\times10^3$ Mpc$^3$ and 460 Mpc$^3$ for sources brighter
than H$_{160}$=26.4, 27.4, 28.4, 29.4, and 30.4, respectively.  Over
35 clusters, if the star formation efficiency and duty cycle remain
fixed, then such a survey should detect 82 $z\simeq 7.5$ LBGs brighter
than J$_{110}$=31.6 (corresponding to a star formation rate of 0.09
M$\rm_\odot yr^{-1}$ at $z\simeq 7.5$ for the source assumptions
discussed in \S 2, Figure \ref{plot_lbg_z7.5}) and 6 $z\simeq 10$
LBGs brighter than H$_{160}$=31.4 (corresponding to a star formation
rate of 0.2 M$\rm_\odot yr^{-1}$ at $z\simeq 10$, Figure
\ref{plot_lbg_z10}).

Adaptive optics (AO) provides the possibility of diffraction-limited
observations from the ground.  If the projected size of the target
objects is small enough, such observations are more efficient than
non-AO observations because the photometric aperture can be decreased,
thereby allowing significantly less noise for nearly the same amount
of flux\footnote{The validity of this statement depends on the strehl
  ratio, which is the ratio of the peak brightness of the stellar
  image to that produced by an ideal optical system. If the strehl is
  very low, then the amount of flux in the diffraction-limited
  aperture will be greatly reduced.}.  One planned survey that aims to
take advantage of AO is the Gemini Genesis Survey (GGS).  This survey
will use a tunable Fabry-Perot  etalon (F2T2, \citealt{Scott06}) on
Gemini with resolution of R=800 to  detect lensed LAEs at $z\simeq
8-10$.  If the projected angular size of the sources is less than
$\simeq$0.03 $\rm arcsec^2$, then GGS should be able to reach a
5$\sigma$ sensitivity of $\rm 3-6\times10^{-18} erg~cm^{-2}~s^{-1}$ in
10 minutes (R. Abraham 2006, personal communication).  Following the
scaling relation derived in \S2, galaxies at $z\simeq 10$ should have
typical sizes of 0.01 $\rm arcsec^2$, small enough to significantly
benefit from AO.  The field of view of F2T2 is
45\arcsec$\times$45\arcsec, ideally suited to imaging the most highly
magnified regions of galaxy clusters.  

We consider a mock GGS survey of 60 clusters; we assume each cluster
is observed for 5 minutes in 40 different wavelength positions between
1.1$\rm \mu m$ and 1.3$\rm \mu m$, allowing the detection of LAEs at
$z=8.1$ to $z=9.7$.  This should take $\simeq 200$ hours of
integration.  A typical cluster provides a magnification gain of $\rm
\times 5, 10, and~30$ over 80\%, 69\%, and 33\% of the F2T2 field of
view \citep{Richard06}.  Taking this as the magnification distribution
for each of the clusters, the total survey volume sensitive to LAEs
brighter than $\rm 10^{41}~erg~s^{-1}$ is 30-86  Mpc$^3$.  The
$z\simeq 8-8.5$ LAE luminosity function (assuming fixed model parameters
from $z=5.7$) suggests that 3-7 sources would be detected at $z\simeq
8-8.5$.  If the effects of supernova feedback are parameterized as in \S2
and \S4, the number density of low luminosity sources is drastically
reduced; in this case, no sources would be detected in the survey.
Alternatively, if all six candidate  LAEs in \cite{Stark07b} are at
high-redshift, then GGS should detect over 30 $z\simeq 8-10$ sources.  

It appears that lensing surveys offer one of the more efficient means
of identifying galaxies at $z\simeq 7-10$ since they are able to
reach sensitivity limits where objects are expected to be much  more
abundant.  However, supernova feedback may drastically reduce the
number of sources detected in these surveys.  Regardless,
spectroscopic confirmation of these sources will continue to remain
challenging until JWST and 20-30 meter ground-based telescopes become
available.

\section{Conclusions}

We have attempted to empirically calibrate the parameters of a  simple
star formation model using observations of star-forming  galaxies
(both LBGs and LAEs) at $z\simeq 6$.  The error budget  used in
fitting the data takes proper account of both the  Poisson and
clustering variance.  We use the calibrated  model to characterize the
physical properties of LBGs and LAEs  at $z\simeq 6$ and extrapolate
it to higher redshifts to make  predictions for upcoming surveys for
galaxies at $z\simeq 7-10$.  Our primary conclusions are as follows:

\noindent {\bf 1.}  We have derived accurate formulae for the
field-to-field variance expected in broadband surveys for LBGs,
narrowband surveys for LAEs, and lensing surveys for LAEs.  For each
survey geometry, there exists a cross-over luminosity $L_c$ below
which the clustering variance dominates over commonly-used Poisson
variance.  In total, the clustering variance accounts for less than
6\% error in narrowband surveys for LAEs in the Subaru Deep Field and
15-20\% error in the $z\simeq 6$ surveys for LBGs in UDF.  The
clustering fluctuations are significantly higher for spectroscopic
lensing surveys reaching up to  100\%.

\noindent {\bf 2.}  LBGs at $z\simeq 6$ are best-fit by a model with a
star formation efficiency of 13\% and a duty cycle of 0.2. The star
formation efficiency suggests that, on average, 87\% of the baryonic
mass of $z\simeq 6$ LBGs still remains in the gas-phase.  The duty
cycle indicates that the current burst of star formation has a
lifetime of 200 Myr, roughly equivalent to the dynamical time of
virialized halos at $z\simeq 6$.  The duty cycle also implies that
80\% of dark matter halos of a given mass are not traced by LBGs.  The
missing halos could have yet to form many stars (due perhaps to
inefficient cooling) or could be quiescent after experiencing a burst
of star formation at earlier times.  This result suggests that the
claimed deficit in ionizing photons from luminous LBGs at $z\simeq 6$
relative to what is required for reionization may be explained by star
formation at higher redshift.

\noindent {\bf 3.} The best-fitting model parameters for LAEs show
some evidence for evolution between $z=5.7$ and $z=6.5$.  However,
most likely  the evolution is not due to a change in the neutral
fraction of the IGM since the parameter that is proportional to the
transmission of Ly$\alpha$ photons through the IGM {\it increases}
between $z=5.7$ and $z=6.5$.  Thus, we consider the evolution to be
tentative because of the large uncertainties in the density of the
lowest luminosity LAEs.  Additional spectroscopic efforts are needed
to improve the spectroscopic completeness at low-luminosities.

\noindent {\bf 4.}  The star formation efficiency of LAEs must be very
low ($\simeq 1$\%) in order to reproduce the low stellar masses
inferred from observations of LAEs at $z\simeq 5$.  Such a low star
formation efficiency is only possible in the context of the models if
the Ly$\alpha$ transmission factor is near unity.  This not only
requires a large escape fraction of Ly$\alpha$ photons from the
galaxy, but also requires that the photons are not substantially
absorbed in their path through the IGM.  The emerging physical picture
is that the LAEs with large EWs are young objects (10-20 Myr) that
have only converted on order 1\% of their baryons to stars (and hence
remain gas-rich) and have not yet produced much dust, allowing a large
escape fraction for Ly$\alpha$ photons.  The intergalactic absorption
decrement in Ly$\alpha$ may be reduced if the ionizing luminosity of
the LAEs or the galaxy groups in which they reside is sufficiently
large so as to strongly ionize the IGM in their vicinity.

\noindent {\bf 5.} We have attempted to fit preliminary data at
$z\simeq 7-10$ by extrapolating our model to higher redshifts assuming
model parameters are fixed at their calibrated $z\simeq 6$ values.
The observed evolution of LBGs between $z\simeq 6$ and $z\simeq 7.5$
can be explained largely by  changes in the host dark matter halos. At
$z\simeq 10$ the picture is more complicated.  Constraints on the
presence of $z\simeq 10$ LBGs in the Hubble UDF \citep{Bou05} are
explained by the hierarchical growth of dark matter halos. In
contrast, the large abundance of $z\simeq 9$ LBGs claimed in the lensing
survey of \cite{Richard06} is difficult to explain without resorting
to a top-heavy IMF or large field-to-field fluctuations.  The
situation is similar for the lensed LAEs.  Two traditional surveys at
$z=8.8$ are consistent with no-evolution in the model parameters from
$z=5.7$, while the abundances inferred from candidate LAEs in a
spectroscopic lensing survey \citep{Stark07b} are only explicable if
the IMF is top-heavy or if the observations are probing an overdensity 
in the underlying mass distribution.

\noindent {\bf 6.}  New instruments that will become available on
ground-based telescopes and HST in the next 2-3 years should greatly
increase the efficiency with which $z\simeq 7-8$ LAEs and LBGs are
detected.  However, unless there is significant upward evolution in the
luminosity function from $z\simeq 6$, detecting $z\simeq 10$ galaxies via
conventional methods will only be feasible if heroic efforts are
undertaken.  With current telescopes, lensing surveys are potentially
better suited to detecting $z\simeq 10$ sources depending on supernova 
feedback.  JWST and thirty-meter
class ground-based telescopes are most likely necessary to detect a
substantial population of objects at $z\simeq 10$.

\noindent {\bf 7.} Constraining reionization at $z\simeq 6-7$ via the
evolution in the LAE luminosity function as probed by future narrowband
surveys will be complicated by clustering and Poisson variance.  Given
the large fluctuations ($\gsim 60$\% for feasible survey geometries and
sensitivities) and small number of sources expected to be detected ($\simeq
7$ assuming no evolution from z=5.7), this technique will only be effective
if the neutral fraction in the IGM evolves very rapidly in the redshift
interval between $z=6.5$ and $z=7.7$.

\subsection*{ACKNOWLEDGEMENTS} 
D.P.S is grateful for the hospitality of the Institute of Theory and 
Computation (ITC) at the Harvard-Smithsonian CfA where this work 
was initiated.   We thank Johan Richard and Rychard Bouwens for 
providing us with the luminosity function from their surveys and Joey 
Munoz for helpful comments on the paper.

\bibliography{journals_apj,mybib}

\newpage
\begin{deluxetable}{lcccccc}
\tablecaption{Variance in High-z Galaxy Surveys}
\tablehead{
\colhead{Field} &  \colhead{Type} & \colhead{Field of View (arcmin$^2$)} &
\colhead{$z$} & \colhead{L$_{Lya,c}$ (erg s$^{-1}$)} & \colhead{L$_{1500}$ (erg s$^{-1}$ Hz$^{-1}$)} 
& \colhead{$\sigma^2_{F2F}(L_c)$}}
\startdata
GOODS   & dropout       & 160  & 6        &    N/A     & 10$^{28.0}$   & 0.002  \\
UDF    & dropout       & 11 & 6        &    N/A     & 10$^{28.2}$   & 0.03    \\
SDF     & narrowband    & 1295 & 5.7      & 10$^{41.7}$ & N/A          & 0.001    \\
Cluster & spec. lensing & 0.13 & 8.5-10.4 & N/A & N/A     & N/A  \\
\enddata
\tablecomments{The field of view for the spectroscopic lensing survey is in the 
source plane, assuming a median magnification of 10.  The crossover luminosity is 
listed as ``N/A'' for the lensing survey because the clustering fluctuations 
are greater than the Poisson noise for all luminosities over which our model predicts 
sources should be detected.
}
\end{deluxetable}

\end{document}